# Small polarons mediated near-room-temperature metal-insulator transition in vanadium dioxide and their hopping dynamics

*Xiongfang Liu[1,#], Tong Yang[2,#], Shanquan Chen[3,#], Jing Wu[4], Chi Sin Tang[1,5,*], Yuanjie Ning[1], Zuhuang Chen[3], Liang Dai[1], Mengxia Sun[1], Mingyao Chen[1], Kun Han[6], Difan Zhou[1], Shengwei Zeng[4], Shuo Sun[1], Sensen Li[7], Ming Yang[2,*], Mark B. H. Breese[5,8], Chuanbing Cai[1], Thirumalai Venkatesan[9], Andrew T. S. Wee[8,10], Xinmao Yin[1,*]*

[1]Shanghai Key Laboratory of High Temperature Superconductors, Institute for Quantum Science and Technology, Department of Physics, Shanghai University, Shanghai 200444, China

[2]Department of Applied Physics, The Hong Kong Polytechnic University, Kowloon, Hong Kong, 100872, China

[3]School of Materials Science and Engineering Harbin Institute of Technology, Shenzhen 518055, China

[4]Institute of Materials Research and Engineering (IMRE), Agency for Science, Technology and Research (A*STAR), 2 Fusionopolis Way, Innovis #08-03, Singapore 138634, Singapore

[5]Singapore Synchrotron Light Source (SSLS), National University of Singapore, Singapore 117603, Singapore

[6]Information Materials and Intelligent Sensing Laboratory of Anhui Province, Institutes of Physical Science and Information Technology, Anhui University, Hefei 230601, China

[7]Department of Electronic Engineering, School of Electronic Science and Engineering, Xiamen University, Xiamen 361005, China

[8]Department of Physics, Faculty of Science, National University of Singapore, Singapore 117542, Singapore

[9]Center for Quantum Research and Technology, University of Oklahoma, Norman, Oklahoma 73019, USA

[10]Centre for Advanced 2D Materials and Graphene Research, National University of Singapore, Singapore 117546, Singapore

#These authors contributed equally to this work.

*Corresponding author: slscst@nus.edu.sg (C.S.T.); kevin.m.yang@polyu.edu.hk (M.Y.); yinxinmao@shu.edu.cn (X.Y.)

**ABSTRACT**

Researchers pursuing advanced photoelectric devices have discovered near room-temperature metal-insulator transitions (MIT) in non-volatile $VO_2$. Despite theoretical investigations suggesting that polaron dynamics mediate the MIT, direct experimental evidence remains scarce. In this study, we present direct evidence of the polaron state in insulating $VO_2$ through high-resolution spectroscopic ellipsometry measurements and first-principles calculations. We illustrate the complementary role of polaron dynamics in facilitating Peierls and Mott transitions, thereby contributing to the MIT processes. Furthermore, our observations and characterizations of conventional metallic and correlated plasmons in the respective phases of the $VO_2$ film offer valuable insights into their electron structures. This investigation enhances comprehension of the MIT mechanism in correlated systems and underscores the roles of polarons, lattice distortions, and electron correlations in facilitating phase transition processes in strongly-correlated systems. Additionally, the detailed detection of small polarons and plasmons serves as inspiration for the development of new device functionalities.

## I. INTRODUCTION

Strongly-correlated electron systems take centerstage in Condensed Matter Physics where the investigation of quantum many-body effects coupled with charge, orbital, spin and lattice degrees of freedom give rise to a rich family of nontrivial critical phenomena such as high-temperature superconductivity (1, 2), charge-density wave order (3, 4) and pair-density wave order (5, 6). Specifically, metal-insulator transition (MIT) phenomenon in Mott insulators emerges as a quintessential feature amongst various quantum mechanical effects (7). The MIT phenomenon can be manipulated via multiple approaches including carrier density modulation, orbital occupancy and photo-induction (8-10). As such, the high-tunable MIT properties enable a diverse range of quantum materials to be utilized in multiple applications related to energy-efficient systems (11), neuromorphic devices (9) and radiative thermal memristors (12). While the prospects of a diverse range of applications have been promising, complex strongly-correlated electronic systems still present a challenge especially in the attempt to identify the underlying physical mechanism governing the MIT processes, which to date, remains to be understood.

Vanadium dioxide ($VO_2$) serves as an ideal platform for the detailed investigation of MIT properties. As an archetypical non-volatile Mott insulator, $VO_2$ has long captivated researchers due to the intriguing MIT process taking place near room temperature ($T \approx 68$ °C). The room-temperature MIT process accords $VO_2$ with potential for practical applications in areas such as the energy sectors, smart windows technology and neuromorphic sensors (9, 11, 13). This first-order phase transition is marked by an abrupt increase in electrical resistivity by several orders of magnitude from a metallic state at high temperature to an insulating state at low temperature. Concomitant with this electronic transition is a structural transition, where the crystal lattice shifts from the high-temperature rutile phase to the low-temperature monoclinic phase. The monoclinic structure is distinguished by the dimerization of vanadium atoms and a tilting relative to the c-axis, a feature that has been closely associated with the insulating state (14). However, the underlying mechanism driving this MIT remains a matter of extensive debate, centering on whether the structural transition or the electronic correlations which takes precedence.

Over the decades, various theoretical frameworks have emerged to explain the MIT process in $VO_2$. Initially, the transition was attributed to a Peierls mechanism, wherein the lattice distortions in the monoclinic phase result in a gap opening at the Fermi level, thereby being stabilized in the insulating phase (15). This perspective was supported by early studies that observed the non-magnetic nature of the monoclinic phase, which is consistent with a typical Peierls insulator (7). However, as more advanced experimental techniques, such as the X-ray absorption spectroscopy and angle-resolved photoemission spectroscopy were used to probe the system (16, 17), it became clear that this lattice-centric view does not fully account for the complexity of the mechanism governing the MIT process.

A competing explanation based on electron-electron interactions, known as the Mott mechanism, posits that strong electronic correlations are the primary driver of the transition, with the lattice distortions playing a secondary role (18). Key experimental findings, such as those by Qazilbash *et al.*, highlighted that in the metallic phase, electron-electron interactions dominate the electronic behavior instead of electron-phonon interactions (19). This suggested that $VO_2$ could undergo a Mott transition, similar to other strongly-correlated electron materials, where insulating behavior emerges due to strong Coulomb repulsion between electrons, independent of any structural phase transition (20, 21). The Mott picture has been bolstered by advances in dynamical mean-field theory (DMFT), which provides a framework for understanding correlated electrons in transition metal oxides (22).

In later years, a growing consensus has emerged that neither the Peierls nor Mott mechanisms alone can comprehensively explain the MIT in $VO_2$. Instead, a cooperative Peierls-Mott transition is increasingly viewed as the most plausible scenario, where both electron-lattice interactions and electron-electron correlations are essential (23, 24). In this hybrid model, the orbital-selective nature of the Mott transition works in tandem with lattice symmetry breaking to stabilize the insulating phase, providing a more holistic understanding of the transition. This model captures the complex coupling between the electronic and structural degrees of freedom, offering insights into the subtle balance of forces governing the phase change in $VO_2$.

Despite these advances, significant challenges remain in fully elucidating the mechanism behind the MIT in $VO_2$. A major limitation lies in disentangling the respective roles of lattice distortions and electronic correlations, as both phenomena are tightly interwoven in the material's phase behavior. With the above consideration, the emergence of polarons (25), a quantum quasiparticle that arises from the coupling between the excess charges and the crystal lattice in strongly-correlated system is the cornerstone to unlock the fundamental understanding of multiple quantum phenomena (26).

The theoretical understanding of polarons began in 1933 with Landau, who suggested that charge carriers could cause local lattice deformation (27). Pekar further formalized this concept by defining the polaron as a quasiparticle resulting from the coupling of a charge with lattice vibrations (28). Two major types of polarons emerged from this work: large polarons, described by Herbert Fröhlich, and small polarons, introduced by Theodore Holstein. Fröhlich's theory applies to polarons with long-range interactions, while Holstein's model focuses on localized, short-range electron-phonon coupling (29).

Over the decades, these theoretical frameworks have expanded, leading to significant developments in both understanding and observing polaron effects. Early models set the foundation for ongoing research, exploring not only single-polaron systems but also complex interactions like bipolarons, magnetic polarons and interfacial polarons (30-33). Experimental breakthroughs, such as optical pump-terahertz probe spectroscopy, Fourier infrared transform spectrometer and soft-X-ray angle-resolved photoelectron spectroscopy (34-36), have further confirmed polaron behavior across a range of materials, including oxides, semimetal, hybrid perovskites and two-dimension interface.

Given the strong influence of charge-lattice coupling, the onset of MIT and its underlying mechanism may fundamentally be mediated by the system's polaron dynamics (30, 37). Besides, polarons are ubiquitous in a diverse range of transition metal oxides (25). Their ability to induce drastic changes, which in turn drives and regulates the corresponding structural and electronic phase transition processes are notable. Notably, extensive computation-led theoretical predictions have been made that point to the prominent existence of small polaron features in $VO_2$ systems, which originate from the short-range coupling between excess charges and the lattice, along with their possible influence on its MIT properties (38-40). Thus, an in-depth investigation on how polaron dynamics would be an effective strategy to unlock the MIT mechanism in multiple condensed matter systems. However, a detailed understanding and direct experimental evidence of MIT effects induced by small polarons in $VO_2$ system remain largely elusive.

Here, we report the observation of a mid-gap state that is present exclusively in the insulating phase of $VO_2$ on $TiO_2$(001) substrate. Based on a comprehensive approach that combines temperature-dependent high-resolution spectroscopic ellipsometry, transport measurements and extensive first-principles studies, we are able to conclusively and convincingly attribute this mid-gap state to the onset of small polarons in insulating phase $VO_2$ and facilitate the analysis of their kinetic behaviors with respect to temperature and structural phases. Moreover, the temperature-dependent optical studies have led to the direct observation of both conventional and correlated plasmons in the metallic and insulating phases of $VO_2$, respectively. The highly sensitive non-destructive photon-in photon-out approach of high-resolution spectroscopic ellipsometry and the full polarization and symmetric features of this optical experimental technique can effectively extract the underlying optical properties of $VO_2$ films and resolve their charge and lattice dynamics. Detailed analyses shows that the onset of small polarons may act as an intermediary to facilitate both the Peierls and Mott phase transitions in regulating the MIT processes of $VO_2$, as depicted in Figure 1a. The identification of a hybrid mechanism of Peierls and Mott transitions synergistically mediating the MIT processes provides a deep understanding of the actual MIT mechanism, not only for $VO_2$, but also for other strongly-correlated oxides, where applications in areas related to photoelectric devices are promising

## II. RESULTS AND DISCUSSION

### A. Validation of Sample Quality

Interfacial strain plays a pivotal role in the behavior of strongly-correlated epitaxial films grown on substrates, influencing key properties such as lattice parameters, crystal orientation, and phase stability. The lattice strain, whether tensile or compressive, arises from lattice mismatches between the film and substrate, which induces significant structural deformations in the epitaxial film. These deformations modulate electron-lattice interactions and orbital hybridization, thereby impacting the material's electronic properties. In such strongly-correlated systems, where electron-electron interactions dominate, interfacial strain can further influence the electronic correlations, modify the spin states, orbital configurations, and ultimately, control the magnetic or electronic phase transition processes.

A well-documented example of such strain effects is seen in $NdNiO_3$, where epitaxial strain induces local distortions of the $NiO_6$ octahedra, mediated by changes in oxygen stoichiometry. These distortions are crucial, as they lead to a small bandgap in the metallic system, thereby driving a metal-to-insulator transition in epitaxial $NdNiO_3$ thin films (41). Similarly, epitaxial strain exerts a profound influence on superconducting transition temperature ($T_c$) in unconventional superconductors such as in $La_{2-x}Sr_xCuO_4$ (LSCO). The oxygen stoichiometry, which governs the doping level and carrier concentration, is highly sensitive to the strain state of the thin film. The kinetics of oxygen exchange at the surface, influenced by both compressive and tensile strain, are crucial in determining the final oxygen content and, consequently, the superconducting properties of LSCO (42). In another example, biaxial strain in $Fe_3GeTe_2$ monolayers enhances ferromagnetism, with the average magnetic moment of Fe atoms increasing monotonically under tensile strain. This effect arises due to strain-induced charge redistribution and the changes in bond lengths, leading to a competition between direct exchange and super-exchange interactions that strengthens the ferromagnetic order (43).

Deciphering the physical properties governed by interfacial strain is essential for uncovering numerous potential applications and guiding future research. The development of strain engineering strategies not only facilitates the design of high-performance energy materials but also enables pioneering functional applications in strongly correlated systems, such as spintronic devices, topological quantum devices, and switchable photovoltaic and neuromorphic systems.

In the case of $VO_2$, interfacial strain not only shifts the MIT temperature but is also related to the formation of intermediate phases, such as the M2 and T phase, which are typically observed during the phase transition processes (44-47). These influences can be attributed to strain-induced changes in orbital occupancy, which enhance electron correlation effects. Consequently, the selection of an appropriate substrate with minimal lattice mismatch is critical for controlling the lattice strain and optimizing the phase transition characteristics of $VO_2$ thin films. Rutile $TiO_2$ emerge as an ideal substrate for $VO_2$ growth due to its compatible lattice parameters and thermal stability. Therefore, the $VO_2$ film was prepared on $TiO_2$ substrate by pulsed laser deposition (PLD) with atomic layer precision and the $VO_2$ film samples are of flat, uniform, ultra-fine morphology at the nanoscale. The thickness of the grown $VO_2$ layer has been measured to be ~28 nm.

X-ray diffraction (XRD) analyses and transport measurements were also conducted on the $VO_2$/$TiO_2$ films to confirm the high crystalline quality of the samples. Figure. 1b displays the XRD pattern of the 28 nm-$VO_2$/ $TiO_2$(001) sample along with the 002-plane for both the rutile $TiO_2$ substrate and the tetragonal-like $VO_2$ film at $2\theta$=62.73 ° and $2\theta$=65.39 °, respectively. The crystal quality of the film was further examined by x-ray rocking curve measurement and atomic force microscopy (AFM) (Supplementary, Section 1.3). Results from these comprehensive XRD analyses indicate the high crystalline quality of the $VO_2$ film where they are of similarly pristine quality as reported in previous studies (48, 49).

Transport measurements are further conducted on the $VO_2/TiO_2(001)$ system (Figure. 1c) where the MIT temperature and temperature-dependent resistivity of $VO_2$ film are once again in very good agreement with previous transport measurements involving similar $VO_2$ film systems (49-52). The resistivity of $VO_2$ film is measured to be $5.9\times10^{-4}$ $\Omega\cdot$cm in the metallic phase at 400 K. The resistivity value increases by four orders of magnitude as temperature decreases into the insulating phase. The resistivity of $VO_2$ film is 9.9 $\Omega\cdot$cm at 250 K. As shown in the inset of Figure. 1c, the $T_{MIT}$ of $VO_2$ film during the heating cycle is 323.3 K. Meanwhile, the MIT takes place at 314.7 K during the cooling process with a small thermal hysteresis width of 8.6 K. The $T_{MIT}$ of $VO_2/TiO_2$ film is lower than bulk $VO_2$ ($T_{MIT}$=341 K) due to the effect of in-plane tensile strain (49). As for the $VO_2$ sample with significantly higher oxygen vacancy concentration, it merely displays a semi-metallic behavior without any onset of the MIT (Supplemental Material, Section 1.3). The $VO_2/TiO_2$ sample with an abrupt transition by four orders of magnitude further affirms that our $VO_2$ sample meet the optical experimental requirements.

**B. Temperature-dependent spectroscopic ellipsometry characterization**

Spectroscopic ellipsometry represents a highly versatile tool for probing and effectively distinguishing between conventional optical band transitions and quasiparticle dynamics that arise due to many-body effects in strongly-correlated materials. It is thus widely utilized to establish the optical response and electron structures in a wide range of condensed matter systems and strongly-correlated materials at the nanoscale (53-55). Detailed optical characterization of quasiparticles was conducted on the $VO_2/TiO_2$ film using spectroscopic ellipsometry at temperatures between 100 K and 400 K where the MIT takes place in the photon energy range between 0.50 and 6.0 eV (Fig. 2). Results of the optical characterization using spectroscopic ellipsometry confirm our transport measurement with prominent changes to the optical parameters caused by the transition process especially with regards to its energy bands and quasiparticle dynamics in both the metallic and insulating states (12, 56). In particular, with the transport measurements confirming the MIT phenomena of the $VO_2$ film during the heating and cooling processes, it behaves as an insulator with a band gap (~ 0.50 eV) below 323.3 K ($T_{MIT}$) (12, 56). It then transforms into a metal at temperatures above 323.3 K (14). Moreover, during the spectroscopic ellipsometry characterization process, we observe the traces of three types of quasiparticles excitations, small polarons, metallic plasmons and correlated plasmons, respectively. The presence of polaron state related to small polarons is detected by analyzing the $\sigma_1$ absorption peaks in the insulating state. Next, metallic plasmons and correlated plasmons are found to distribute in the loss-function (LF) spectra. The properties of small polarons and plasmons are discussed in detail below, respectively.

**The observation of polaron state**

The optical conductivity of the $VO_2$ films (Figure 2a) were obtained from a combination of high-resolution spectroscopic ellipsometry measurements (0.5−6.0 eV) and transport measurements (0 eV). We plot the spectral weight $SW = \int_{w_2}^{w_1} \sigma_1(\omega')d\omega'$ in Figure 2b for $VO_2$. An enhancement of in the spectral weight of $\sigma_1(\omega)$ below 1.72 eV comes from spectral weight above 1.72 eV in metallic state. This is a contribution from the Drude response of free electrons in low-energy region (19, 57). Based on spectral-weight transfer and optical features, $\sigma_1(\omega)$ is divided into five photon energy regions: I (0-1.62 eV), II (1.62-3.28 eV), III (3.28-4.10 eV), IV (4.10-4.72 eV) and V (4.72-6.0 eV). As temperature decreases, there is an increase of spectral weight in regions II, III, IV, V and decrease of spectral weigh in region I can be clearly seen in Figure 2c. The changes of the number of effective carriers per formula unit $N_{eff}(\omega)$ for different spectral regions present the same tendency with the changes of spectral weight (Supplementary, Section 2.3), as shown in Figure 2d. The changes of spectral weight and the number of the effective carriers are consistent with the conductivity behavior, indicating the occurrence of the MIT process. This is also an important sign of strong electronic correlations (58).

Figures. 3a-b display the optical conductivity, $\sigma_1$, in the metallic ($T$=400 K) and the insulating states ($T$=100 K) elucidated from high-resolution spectroscopic ellipsometry measurements (0.5−6.0 eV). In the metallic phase at 400 K (Figure 3a), three prominent features, labelled α (~ 2.75 eV), β (~ 3.38 eV) and γ (~ 5.07 eV), can be observed. To derive a more accurate set of information on the energy position and shape of the respective features, the $\sigma_1$ spectra is then modelled using Lorentzian peaks ($L_\alpha$, $L_\beta$, $L_\gamma$) and a low-energy Drude feature based on the Drude-Lorentz model [(Figure 3a); Supplementary, Section 2.4]. The broad Drude response is a clear indication of the $VO_2$'s metallic properties in its metallic phase (54). Based on the analyses of the $\sigma_1$ spectra elucidated from the spectroscopic ellipsometry measurements (Supplementary, Section 2.5), the weak shoulder feature α (~ 2.75 eV) can be attributed to the optical transition from the filled O 2$p$ band to the half-filled $d_{||}$ band, while the feature β (~ 3.38 eV) is ascribed to the optical transition from the filled O 2$p$ band to the half-filled $\pi^*$ band. Meanwhile, feature γ (~5.07 eV) is attributed to the optical transition between half-filled $d_{||}$ band and empty $\sigma^*$ band. The positions and origins of the respective optical features and the lineshape of the $\sigma_1$ spectrum is in good agreement with previous studies (57, 59).

With the metal-to-insulator transition of $VO_2$ taking place at 323.3 K with the almost-simultaneous onset of Mott and Peierls transition (7, 16, 23, 60), it results in very significant changes to both the electrical conductivity and lattice structure of the $VO_2$ system. The Lower Hubbard band (LHB) and Upper Hubbard band (UHB) are two subbands split by half-filled Hubbard model due to the influence of Hubbard intra-atomic Coulomb repulsion energy (57). LHB is a fully-filled energy band located below the Fermi level while the UHB is an empty energy band located above the Fermi level (61). This results in a corresponding change to the $\sigma_1$ spectra below 323.3 K where $VO_2$ is now in the insulating phase. The insulating phase displays a sizable energy gap of ~0.50 eV at 100 K —in agreement with the previous report for the insulating phase of $VO_2$ (57). At 100 K, several distinct optical features in the $\sigma_1$ spectrum can be observed (Figure 3b). Similar to the metallic phase $VO_2$, $\sigma_1$ spectrum is modelled using the Lorentzian oscillators, in this case, using $L_A$, $L_B$, $L_C$, $L_D$, $L_E$. The feature A at ~ 1.26 eV is attributed to the optical transition between LHB (Lower Hubbard band) and $\pi^*$ band. Feature C at ~3.39 eV in the insulating phase can be attributed to the interband transition between the LHB and UHB (Upper Hubbard band). Meanwhile, the hump feature E (~5.33 eV) can be accounted for by the optical transition from O 2$p$ band to $\pi^*$ band and from LHB to $\sigma^*$ band (59).

In addition to the previously established optical absorption features as described above, two other previously unidentified features (labelled B and D) located at ~ 3.02 eV and ~ 4.26 eV, respectively, have also been observed (Figure 3b). Intriguingly, the spectral weight of features B and D are relatively low, while their energy positions do not shift with varying temperature (Supplementary, Section 3.4). Eventually, features B and D disappear above $T_{MIT}$ along with the formation of a Drude response in the metallic phase. Based on the energy positions and the relatively small spectral weight of features B and D, it can be deduced that they are attributed to the formation of a new narrow mid-gap state (62). Thus, features B and D are optical transitions from this mid-gap state to unoccupied band.

In addition, since the low SW of $\sigma_1$ in the low photon energy region (<2.0 eV) in the insulating state, it is difficult to accurately fit the $\sigma_1$ by using only these Lorentzian oscillators. A more in-depth analysis of the spectroscopic ellipsometry data is conducted and they are plotted in the form of a differential spectra of optical conductivity between various temperatures and 300 K in insulating state, $\Delta\sigma_1(\omega, T)=\sigma_1(\omega, T)- \sigma_1(\omega, 300\ K)$, as shown in Figure 3c. An independent feature can be observed at ~0.7 eV (labelled Δ) across different temperatures (Supplementary, Section 2.6). This feature may be attributed to the optical transition from the mid-gap state to the π* band. As discussed in detailed thereafter, we attribute this narrow mid-gap state to the polaron state that is present in $VO_2/TiO_2$ film.

The inevitable presence of oxygen vacancies (63) in $VO_2$ film has provided the necessary conditions in the formation of small polarons via the coupling between excess electrons and V-lattice sites. Meanwhile, in-plane tensile strain of $VO_2$ film increase the hopping energy of small polaron, which is beneficial for stabilizing the small polaron(32, 49). It has also been reported that the transport properties of insulating state $VO_2$ is dictated by the small polaron hopping dynamics, and these hopping dynamics is in turn determined by the amplitude of the thermal lattice vibrations (64, 65). Hence, we utilized the small polaron hopping model (Supplementary, Section 3.2) to model and confirm the presence of the small polarons in insulating phase $VO_2$. The transport curve displayed in Figure 3d shows that it is in good agreement with the theoretical small polaron hopping model in the insulating phase below $T_{MIT}$ (323.3 K). In the metallic state above $T_{MIT}$, the transport data begins to deviate from the theoretical model. This observation once again provides evidence that while small polarons are present in insulating phase $VO_2$, they begin to disappear in metallic state $VO_2$ above $T_{MIT}$ due to the increase in thermal lattice vibration.

Having provided the experimental evidence that small polarons are present in the insulating state $VO_2$ based on a detailed analysis of the transport data below and above the $T_{MIT}$, it is essential to clearly attribute features B and D observed in the optical spectra (Figure 3b) to the presence of the polaron state in insulating phase $VO_2$. A previous theoretical study has indicated that polaron state is formed at energy region in the order of ~1 eV below the conduction band minimum (CBM) in large bandgap oxide systems (25). With the relatively small bandgap of insulating $VO_2$, not only is the formation of the small polaron state at ~0.6 eV below the empty $\pi^*$ band at 100 K (as discussed later in Figure 4b) theoretically consistent, but it also agrees with experimental detected polaron state in other small bandgap oxide system (66). Hence, by comparing evidences from our optical studies with previous theoretical studies detailing the presence of the polaron state in insulating phase $VO_2$, they provide substantial evidences that a small polaron state is present below the Fermi level with features B (~ 3.39 eV) and D (~ 4.26 eV) in the $\sigma_1$ spectra (Figure 3b) being the optical transitions from the polaron state to the UHB and $\sigma^*$ band, respectively. While our experimental studies have provided very conclusive evidences pointing to the presence of small polarons in insulating phase $VO_2$, further confirmation of this quantum quasiparticle will be made via a set of computational study as discussed thereafter.

To further substantiate the experimental evidences that the mid-gap state is attributed to the formation of small polarons, detailed First-principles calculations have been performed to scrutinize the effects of electron-phonon coupling in $VO_2$ (Supplementary, Section 4). When excess electrons are introduced to the non-polaron $VO_2$ lattice (Figure 4a), these excess electrons are redistributed in the lattice where there is a tendency that they become self-trapped around vanadium sites (Figure 4b), which in turn induces the formation of the polaron state. Meanwhile, these localized electrons can also be found in neighboring vanadium sites in the vicinity where the polaron states are formed (67). In addition, the system in the non-polaron state (Figure 4a) is 0.771 eV higher in energy than the polaron state (Figure 4b). This is a clear indication that it is energetically more favorable for the excess electrons to be localized in the V lattice sites to form small polarons than in the non-polaron state where the electrons are homogeneously distributed across the lattice system. In the specific scenario where the excess charges are localized in the form displayed in Figure 4c (magnified image in Figure 4d) where small polarons are formed in the $VO_2$ lattice, the shape of the partial charge density implies that the excess electron mainly occupies the V $3d_{yz}$ and $3d_{xz}$ orbitals (Figure 4d). As compared to the non-polaron state counterpart, while the charge trapping at the V lattice site further increases the long V-V distance, the short V-V distance remains virtually unchanged. This is likely due to the aforementioned charge transfer from that neighboring V site.

Further comparison is made between the visualized partial charge density of state (PDOS) of the non-polaron state as shown in Figure 4e corresponding to the homogeneous excess charge distribution of Figure 4a and the PDOS in which self-trapping of charges takes place (Figure 4f) as depicted in Figure 3b, where polaron states are formed. The PDOS of the non-polaron state $VO_2$ (Figure 4e) based on the GGA+U level of theory. The calculated band gap is 0.601 eV—in agreement with what is previously reported in insulating phase $VO_2$ (12, 56). As for the visualized PDOS where small polarons are present (Figure 4f), it is interesting to note that a mid-gap state (marked by arrow) corresponding to the polaron state can now be observed. This polaron state emerged below the Fermi level, and it mainly consists of the V$t_{2g}$ states. Very importantly, the results of these DFT studies show clear consistency with our experimental findings. Specifically, the mid-gap state that emerged in the visualized PDOS can be attributed to the formation of the polaron state. This provides compelling evidence to substantiate our experimental results for the presence of small polarons in insulating state $VO_2$.

**The observation of plasmon excitations**

To further analyze the quasiparticle features and opto-electronic properties of the $VO_2$ system and distinguish between the insulating and metallic phases, the optical loss-function (LF) is further derived with the spectra displayed in Figures 5 (Supplementary, Section 2.7). Metallic phase $VO_2$ shows clear metallic behavior with the prominent Drude response in $\sigma_1$ (Figure 3b) and zero-crossing of $\varepsilon_1$ (Figure 5a) at ~ 1.12 eV at 400 K. While the coincidence of the zero-crossing in the $\varepsilon_1$ spectra and the LF peak $A_1$ (Figure 5b) at the same photon energy position marks the presence of metallic plasmons in metallic phase $VO_2$. Similar observations of metallic plasmons have also been reported in previous literatures (68, 69). An essential observation is depicted in the $\varepsilon_2$ and LF spectra of $VO_2$ (Figures 5c-d). In the LF spectra of metallic phases (400 K) and insulating state (100 K), several features, namely $B_1$, $B_2$, $B_3$ and $C_1$, $C_2$, $C_3$, $C_4$, $C_5$ correspond one-to-one with the features in $\varepsilon_2$ spectra: $B_1'$, $B_2'$, $B_3'$ and $C_1'$, $C_2'$, $C_3'$, $C_4'$, $C_5'$, respectively. These observations suggest that the simultaneous presence of the features in both the $\varepsilon_2$ and LF spectra where they indicate a coupling between optical and plasmonic excitations in metallic and insulating $VO_2$ (70) (Supplementary, Section 2.8). Therefore, these corresponding features in both the metallic and insulating phases of $VO_2$ can be regarded as a form of correlated plasmons, arising from the collective excitations of correlated electrons. These collective features are signatures of correlated plasmonic excitation where they arise due to the presence of strong electron−electron and electron−hole interactions in the $VO_2$ film (71, 72).

While plasmon dynamics in the respective phases of $VO_2$ do not display any direct association with the system's MIT processes, the observation and characterization of both conventional (metallic) and correlated plasmons in the distinct phases of the $VO_2$ film offer valuable insights into the evolution of collective charge dynamics within the system as it undergoes transformations between structural and electronic phases. Furthermore, a comprehensive grasp of plasma properties in $VO_2$ is crucial for their utilization in optoelectronic devices.

**C. Potential mechanism relating small polarons and the MIT processes**

Based on the energy positions of the major features in Figures 3a-c and combining it with previous X-ray Absorption Spectroscopic (XAS) characterization studies on the energy band and corresponding electronic structures of $VO_2$ (Supplementary, Section 3.4), an energy band schematic of the $VO_2$ system in both the metallic (400 K) and insulating phase (100 K) can be plotted as shown in Figures 6a-b, respectively (8, 16). Specifically, the polaron state emerges as a distinct feature that is located between the filled LHB and the empty $\pi^*$ band in insulating phase $VO_2$ (Figure 6b).

Having confirmed the presence of small polarons and by analyzing the temperature-dependent energy band properties, this provides a new perspective on how small polarons facilitate the MIT processes through the opening of the bandgap in insulating $VO_2$ film which is yet to be fully understood. A pictorial representation of the electron dynamics above and below $T_{MIT}$ has been provided in Figures 6c-d, respectively. Specifically, Figure 6c depicts the electron dynamics in the metallic phase at high temperature where they manifest themselves in the form of free and correlated electrons, while Figure 6d shows the localization of the electrons due to the gradual enhancement of electron-phonon interaction with decreasing temperature. In the high-temperature metallic phase where the system is significantly more energetic due to the considerable presence of thermal energy, the effects of thermal lattice vibration is significant. These high-temperature thermal effects dominate the system’s lattice dynamics and along with the dynamics of the free electrons, they outweigh the effect of electron localization (Figure 6c) (73). Meanwhile, in this high-temperature metallic phase, the excitations of free and correlated electrons that manifest themselves in the form of metallic and correlated plasmons further prevent the localization of the electron due to their energetic periodic motion (Figure 6c). As a result, the collective effects of such high-temperature electron dynamics prevent the formation of small polarons. However, as temperature decreases and as it approaches $T_{MIT}$, there is a corresponding weakening of the thermal lattice vibrations which in turn result in the gradual enhancement of electron-phonon interaction and, at the same time, the trapping of free electrons at the V-lattice sites and in their vicinity (Figure 6d) (19, 74). This process of electron trapping due to the progressive amplification of the electron-phonon coupling greatly facilitates the formation of small polarons which is accompanied by a series of lattice distortion and symmetry breaking (30). With the increase in electron-phonon interactions, there is an amplification in lattice instability and that such structural distortions are quintessential conditions that lead to the phase transition process in the form of a Peierls transition (23, 24, 75). Fundamentally, this phase transition process is driven by the structural instability, and in order to stabilize the once again, an energy gap opening takes place at the Fermi surface due to the formation of a charge density wave. The formation of small polarons essentially facilitates the onset of Peierls transition via the localization of charge carriers and the distortion the lattice structure, thereby driving the instability and leading to the phase transition.

At the same time, strong electron-electron interactions play a significant role in driving the metal to an insulator transition as described by the Mott theory where the competition between electron kinetic energy (charge delocalization) and Coulombic repulsion (charge localization) determines the phase of the system (76). This transition phase transition process typically takes place in systems such as transition metal oxides where electron correlations are significant (77, 78). In systems where strong electron correlations are predominant, the formation of small polarons due to electron-phonon coupling enhances the localization of charge carriers, making it easier for a system to transition to an insulating state, as required by the Mott criterion (79). In addition, the formation of small polarons can enhance the effective Coulomb repulsion between electrons. This aids in promoting the Mott insulating phase, where the electron-electron (Coulombic) repulsion dominates over kinetic energy.

In the case of $VO_2$ which specifically describes the dynamics of electrons in the V 3*d* channels (80), there is a significant reduction in free electrons as they become localized at the V lattice sites to form small polarons. The reduction of free electrons significantly diminishes the electronic screening effect and this reinforces the system's electron correlations especially at low temperature (19, 81). In addition, the induced localization of the electrons reduces the bandwidth of the electronic states. Consequently, the augmented electronic correlation establishes a conducive condition where it tips the balance in the competition between electron kinetic energy (charge delocalization) and Coulombic repulsion (charge localization) in favor of the insulating Mott phase.

Through the localization of electrons and the induction of lattice distortions in the formation of small polarons, the significant role that small polarons can play in mediating both the Peierls structural transition and the Mott electronic phase transition processes is evident. It can therefore be deduced that small polarons serve as an important intermediary in facilitating a hybrid mechanism comprising the aforementioned phase transitions in $VO_2$.

## Ⅲ. OUTLOOK

Currently, the cooperative Peierls-Mott model represents the most comprehensive explanation of the MIT in $VO_2$. Further in-depth research in the field of $VO_2$ is likely to benefit from advances in ultrafast spectroscopy and nanoscale imaging techniques (10, 14). These will offer the potential to probe the MIT with unprecedented resolution. These methods could provide direct observations of the coupling between electronic and structural changes, allowing researchers to better understand the temporal sequence of events during the transition. Meanwhile, first-principles calculations and improved DMFT models are expected to offer deeper insights into the role of electronic correlations and their interaction with lattice distortions, potentially resolving long-standing ambiguities regarding the relative importance of the Peierls and Mott mechanisms (40, 82).

A promising direction for future development in this discipline involves the exploration of other diverse methods to control and tune the MIT in $VO_2$ via external stimuli, such as strain, doping, or electric fields. Such approaches could enable the manipulation of the transition temperature and the electronic properties of $VO_2$, paving the way for its integration into novel electronic and photonic devices. Applications in ultrafast switches, memory devices, and neuromorphic computing have already been proposed (9, 10, 83), and further advances in the understanding of the MIT could accelerate the realization of these technologies. The ability to modulate the MIT through precise control of external parameters represents a crucial step toward harnessing the unique properties of $VO_2$ for practical applications, while also offering a rich platform for exploring the fundamental physics of correlated electron systems.

Beyond the domain of the $VO_2$ system and the mechanisms underlying the MIT process, the development of polaron research presents new prospects and opportunities where they encompass a diverse variety of materials and phenomena. Polarons play a crucial role in determining properties such as electrical conductivity, surface reactivity, and magnetoresistance in various materials, including transition-metal oxides, conducting polymers, and 2D materials (25). Computational tools such as density functional theory (DFT) and dynamical mean-field theory (DMFT) have enabled more precise calculations of polaron properties, such as binding energy, effective mass, and mobility. Despite these advances, theoretical models like Landau-Pekar and Fröhlich's pose considerable limits in describing real materials in an effective way. This is because many theoretical models oversimplify the electron-phonon interactions and are unable to account for the dynamics in non-polar systems (84, 85). Current challenges include the need for models that capture both short- and long-range coupling and the accurate representation of polaron behavior in hybrid materials like halide perovskites, where complex electron-phonon interactions complicate the picture.

Looking ahead, advancing polaron research will require more sophisticated models and computational techniques that can better integrate first-principles calculations with realistic material properties. One key area of focus will be the development of unified frameworks that address both short-range and long-range electron-phonon interactions. Additionally, research into low-dimensional systems, such as 2D materials, is expected to yield new and intriguing insights into how polarons behave under spatial confinement. Polaron studies will also expand into multi-polaron systems, where interactions between multiple polarons may lead to novel phenomena such as bipolaron formation, with implications for high-temperature superconductivity. The role of polarons at material interfaces, especially in van der Waals heterostructures, will be another area of intense study, potentially opening new paths for electronic and optoelectronic devices.

Polaronic effects hold promise for a wide range of applications. In thermoelectric materials, controlling polaron mobility could enhance energy efficiency, while in organic semiconductors, understanding polaron dynamics is crucial for optimizing charge transport. Hybrid perovskite solar cells, where polarons influence both light absorption and charge separation, represent another key application area. In catalysis, the role of polarons in mediating surface reactions could be harnessed to improve catalytic performance. Beyond these practical applications, polarons are also relevant in emerging quantum technologies, where they could be used to manipulate charge and spin states in quantum dots and other nanoscale devices. This potential for applications in quantum computing and memory devices makes polaron research a vital field for future technological advancements.

## IV. CONCLUSION

In conclusion, we have clearly and convincingly demonstrated the presence of small polarons in insulating phase $VO_2$ film through a comprehensive study that combines a range of experimental and first-principles investigation. This study highlights the small polarons intimately related phase transition processes underscore the assertion that both electron-phonon and electron-electron interactions may play crucial roles in driving the metal-insulator transition processes in $VO_2$ as well as in other strongly-correlated systems. Further experimental observations have also shown the presence of conventional and correlated plasmons in $VO_2$ system. Meanwhile, these new insights in strongly-correlated oxides allow for potential applications related to energy conversion, such as thermoelectric generators, neuromorphic sensors and photoelectric devices. Hence, regulating quasiparticles generation and dynamics is undoubtedly advantageous to further device applications.

## SUPPLEMENTARY MATERIAL

Discussion on sample preparation (PLD method); crystalline quality characterizations (XRD, X-ray rocking curve, transport measurements); Spectroscopic Ellipsometry measurements (principle, methods, model, and data analysis); detection of quasiparticles (analysis of small polarons, metallic plasmons, and correlated plasmons; illustration of electrical band structure and theoretical model fitting); details of first-principles calculations; supplementary figures and associated references (PDF).

## AUTHOR CONTRIBUTIONS

X.L., T.Y., S.C. contributed equally to this work. X.Y. conceived the project. K.H., S.Z., T.V., S.C. synthesized the samples and K.H., S.Z. performed the electron transport experiments and analyzed the data, C.S.T. performed the XRD experiments and analyzed the data and X.L., C.S.T., M.S., M.C., S.S.L. performed the SE experiments and analyzed the data. T.Y., M.Y. performed the DFT calculations. X.L. and C.S.T. wrote the manuscript, with input from all the authors.

## Conflict of Interest

The authors have no conflicts to disclose.

## ACKNOWLEDGMENTS

This work was supported by National Natural Science Foundation of China (Grant No. 52172271, 12374378, 52307026), the National Key R&D Program of China (Grant No. 2022YFE03150200), Shanghai Science and Technology Innovation Program (Grant No. 22511100200, 23511101600), the Strategic Priority Research Program of the Chinese Academy of Sciences (Grant No. XDB25000000). J.W. acknowledges the Advanced Manufacturing and Engineering Young Individual Research Grant (AME YIRG Grant No. A2084c0170) and the SERC Central Research Fund (CRF). C.S.T acknowledges the support from the NUS Emerging Scientist Fellowship. S.S.L. would like to thank the funding support from National Natural Science Foundation of China (NSFC) (No. 62175206). M.Y. acknowledges the funding support from The Hong Kong Polytechnic University (project number: 1-BE47, ZE0C, ZE2F and ZE2X).

### DATA AVAILABILITY

The data that support the findings of this study are available within the article and its supplementary material.

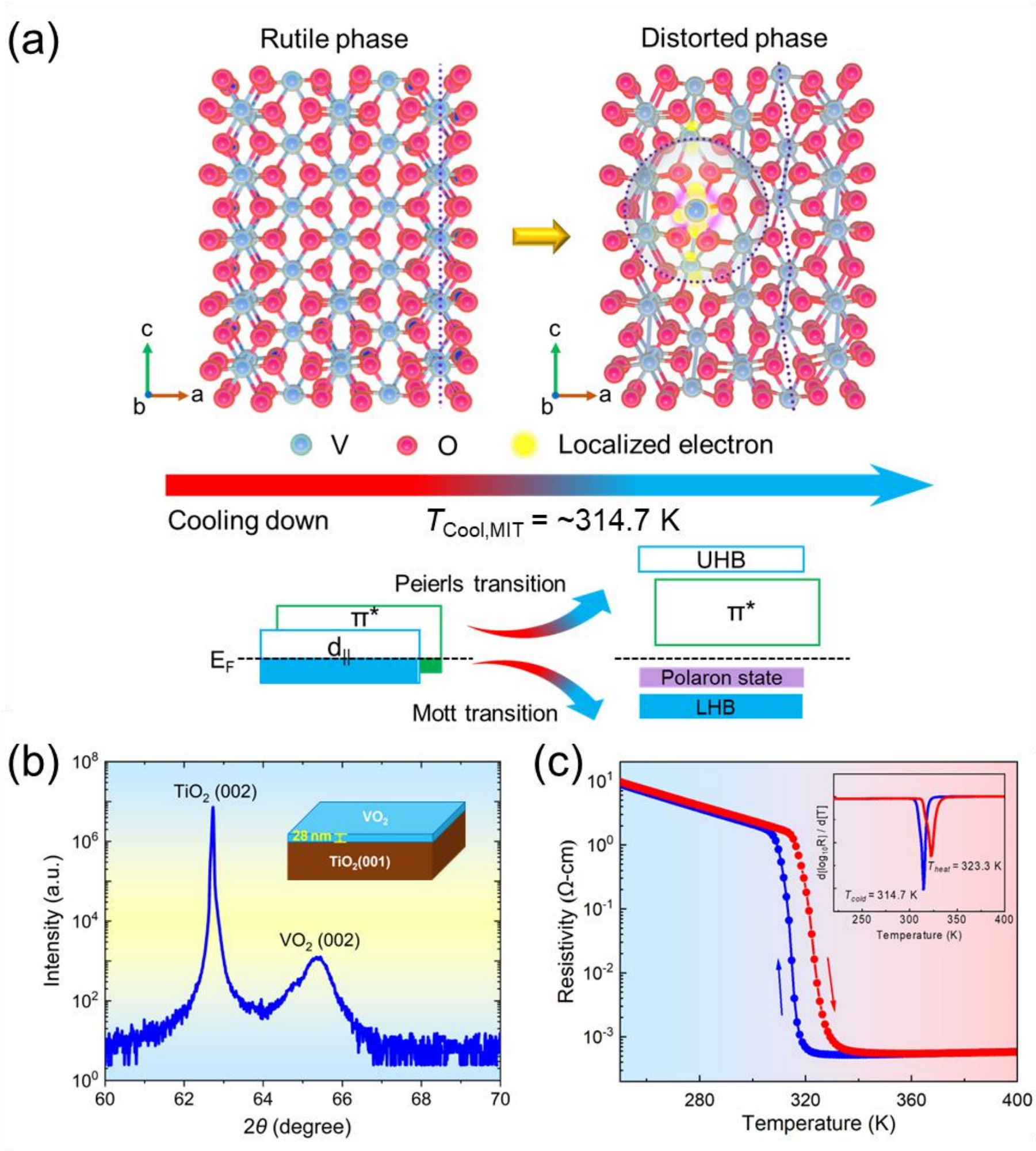

**Figure 1**. The intermediary role that small polarons play in the MIT processes of $VO_2/TiO_2$ film and results from the XRD and transport measurement on the sample. (a) Schematic that depicts the localized electrons in the vicinity of a distorted $VO_2$ lattice which is indicative that small polarons are present in insulating phase $VO_2$. Presence of small polarons may possibly mediate the onset of Peierls and Mott phase transitions near $T_{\mathrm{cool,\ MIT}}$ (314.7 K) during the cooling processes. These physical terms $E_F$ (Fermi level), $\pi^*$, $d_{||}$, UHB (Upper Hubbard Band) and LHB (Lower Hubbard Band) present different band structure of $VO_2$ system. Peierls transition and Mott transition are two types involving MIT processes. (b) Structural analysis of epitaxial $VO_2$ film grown on $TiO_2(001)$ substrate based on XRD characterization. $\theta$-$2\theta$ scan around the $TiO_2(002)$ and $VO_2(002)$ peak where no other peaks have been found. (c) Temperature-dependent resistivity curves of $VO_2$ films grown on $TiO_2(001)$ substrate. The inset shows the corresponding $d[\log_{10}(R)]/d[T]$ curves.

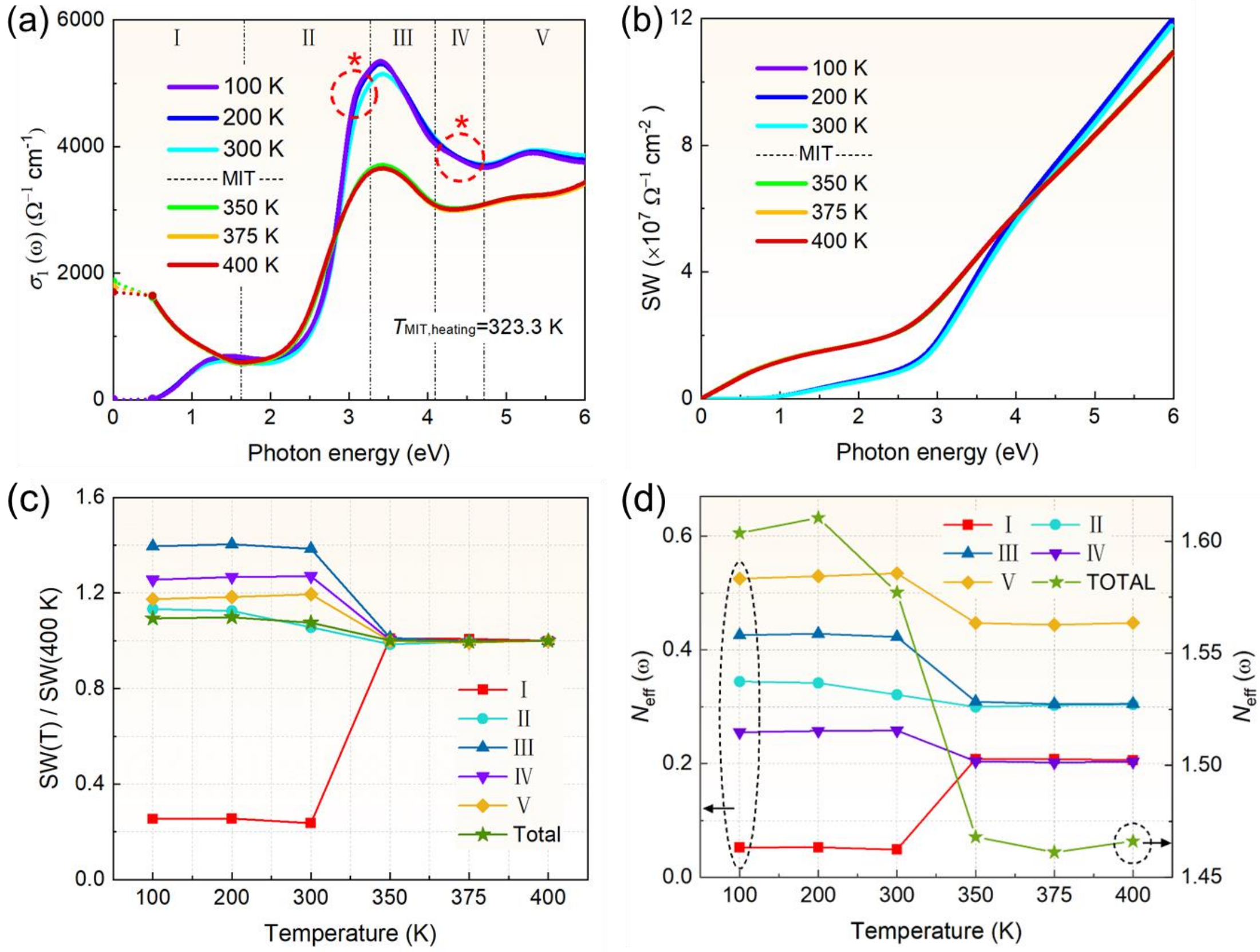

**Figure 2.** Optical conductivity, $\sigma_1$, spectra of the $VO_2/TiO_2$ film. (a) $\sigma_1$ spectra of $VO_2/TiO_2(001)$ film from 0 to 6.0 eV at the respective temperatures in both its insulating and metallic phases (where $T_{\mathrm{MIT}}$ = ~323.3 K). The $\sigma_1(0)$ is obtained from transport measurements and the linear interpolation is used to estimate the $\sigma_1$ from 0 to 0.50 eV. (b) The spectral weight SW is plotted by integrating the $\sigma_1(\omega)$. (c) Ratio of the spectral weight defined as SW(T)/SW(400 K) and (d) The number of the effective carriers per formula unit $[N_{eff}(\omega)]$ as a function of temperature in different spectral regions: I (0-1.62 eV), II (1.62-3.28 eV), III (3.28-4.10 eV), IV (4.10-4.72 eV), V (4.72-6.0 eV) and total (0-6.0 eV). Temperature is described as T (K).

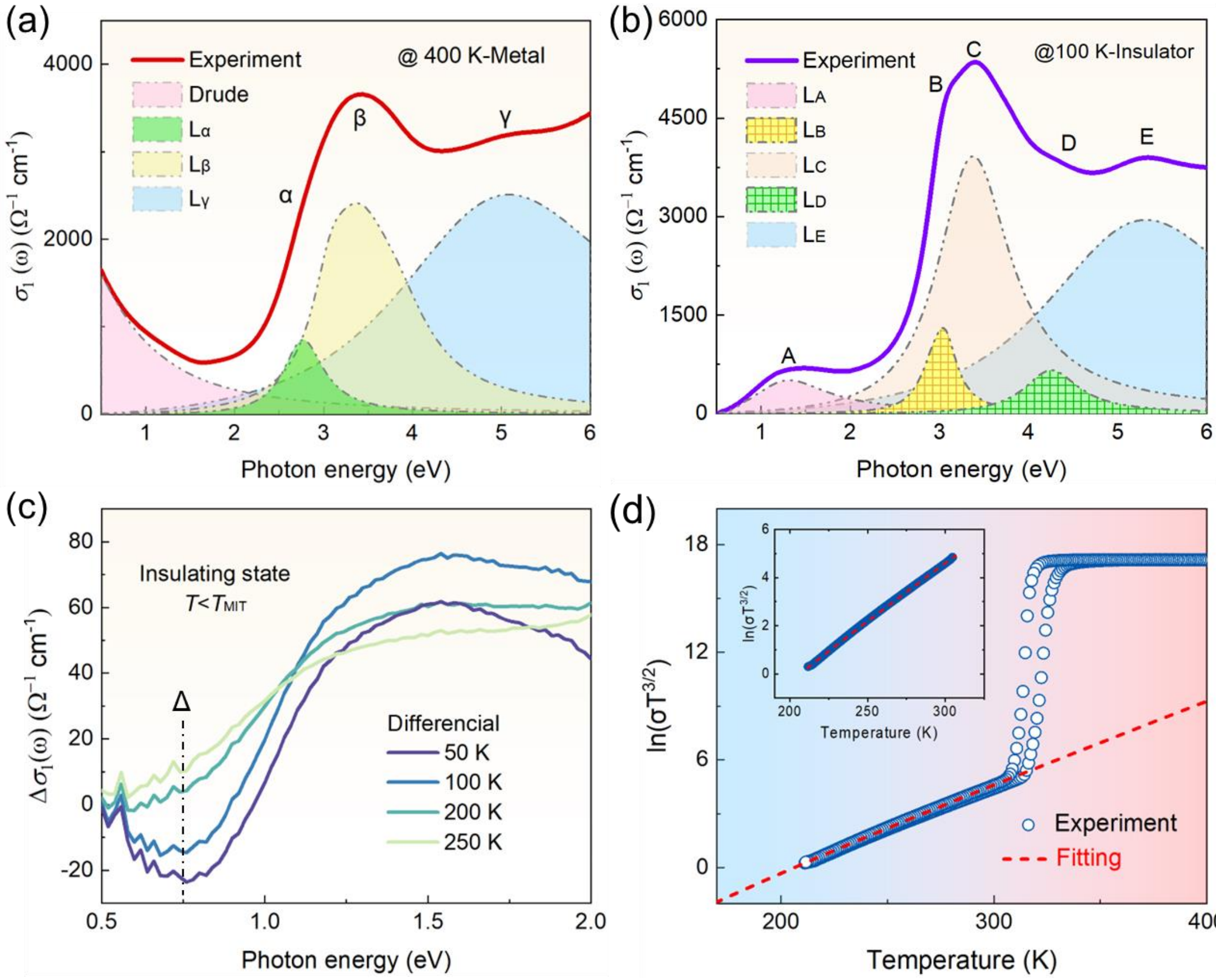

**Figure 3.** Optical feature fitting and transport measurements of the $VO_2/TiO_2$ film. (a-b) $\sigma_1$ spectra of $VO_2$ and their respective fitting curves in (a) metallic phase $VO_2$ film at 400 K, and (b) insulating phase $VO_2$ film at 100 K to account for the respective peak features present in the spectra. The solid curve represents the actual $\sigma_1$ spectra while the peaks in their respective colors represent the fitted curves. A Drude feature has been added to metallic phase $VO_2$. (c) The differential spectra of the optical conductivity $\Delta\sigma_1(\omega, T)=\sigma_1(\omega, T)-\sigma_1(\omega, 300\text{ K})$ for $VO_2/TiO_2$ film,. The symbol 'Δ' (~0.7 eV) denotes an optical feature detected in the $\Delta\sigma_1(\omega, T)$ spectrum. (d) ln $(\sigma T^{3/2})$ vs $T$ curves. The red dash line is fitted to the experimental data using the hopping conductivity of small polarons. Inset: ln $(\sigma T^{3/2})$ vs $T$ below $T_{MIT}$. Indicating that the activated behavior of small polarons only takes place in the insulating phase below $T_{MIT}$. Data shows very good agreement with the theoretical fitting line.

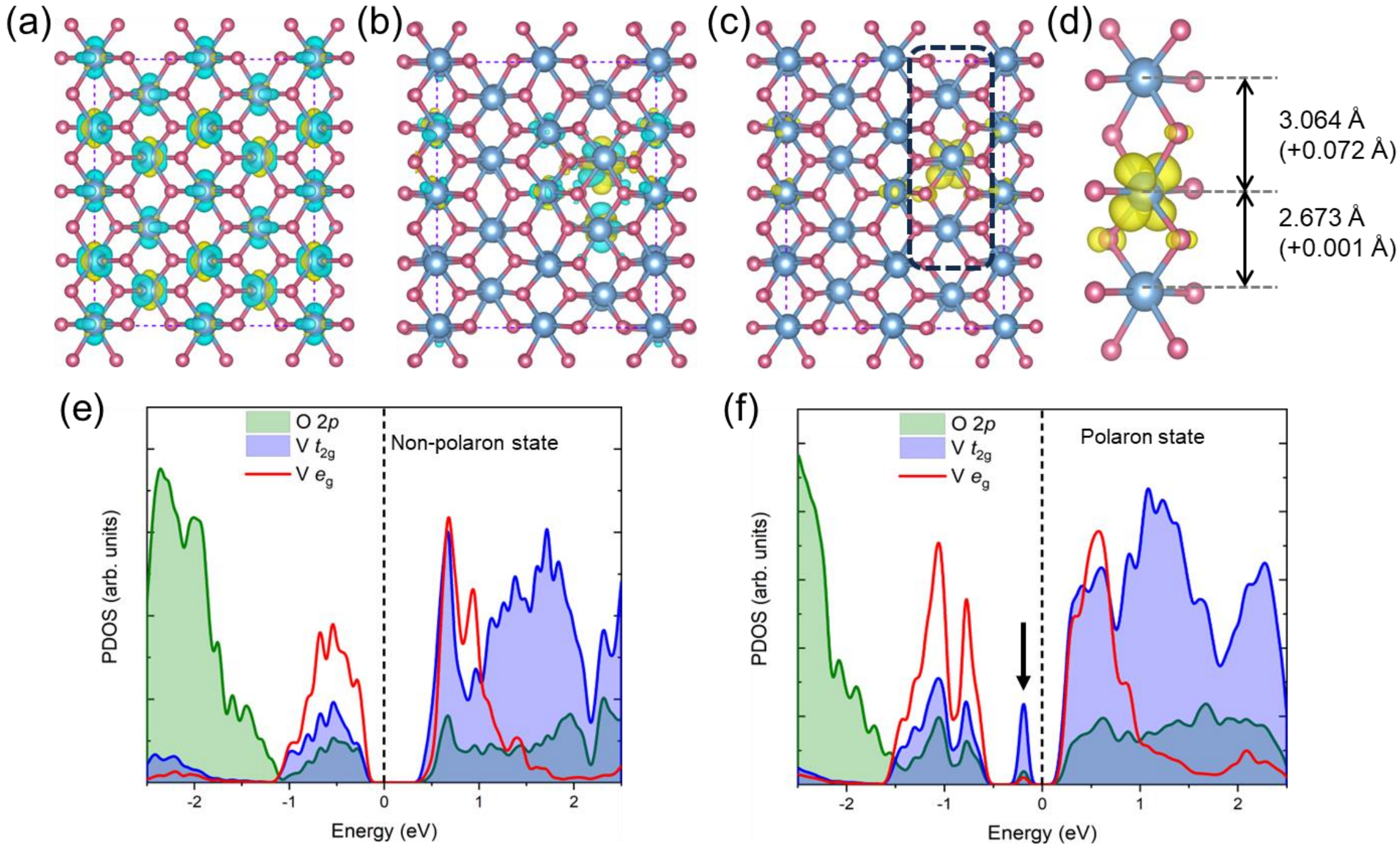

**Figure 4.** First-principles calculations and plasmon excitations spectra of the $VO_2/TiO_2$ film. Excess charge density distribution in the (a) non-polaron state and (b) polaron state. The charge accumulation/depletion is denoted by the yellow/cyan iso-surfaces with an iso-value of 0.0015 e/Å$^3$ (non-polaron state) and 0.007 e/Å$^3$ (polaron state), respectively. (c) Visualized partial charge density associated with the mid-gap state described thereafter in (f). (d) Magnification of the localized charges enclosed in the black dashed rectangle of (c) with the iso-value is set to 0.007 e/Å$^3$. (e) Projected density of states (PDOS) of non-polaron $VO_2$ and (f) PDOS of $VO_2$ in the polaron state where the Fermi level has been set to 0 in both scenarios.

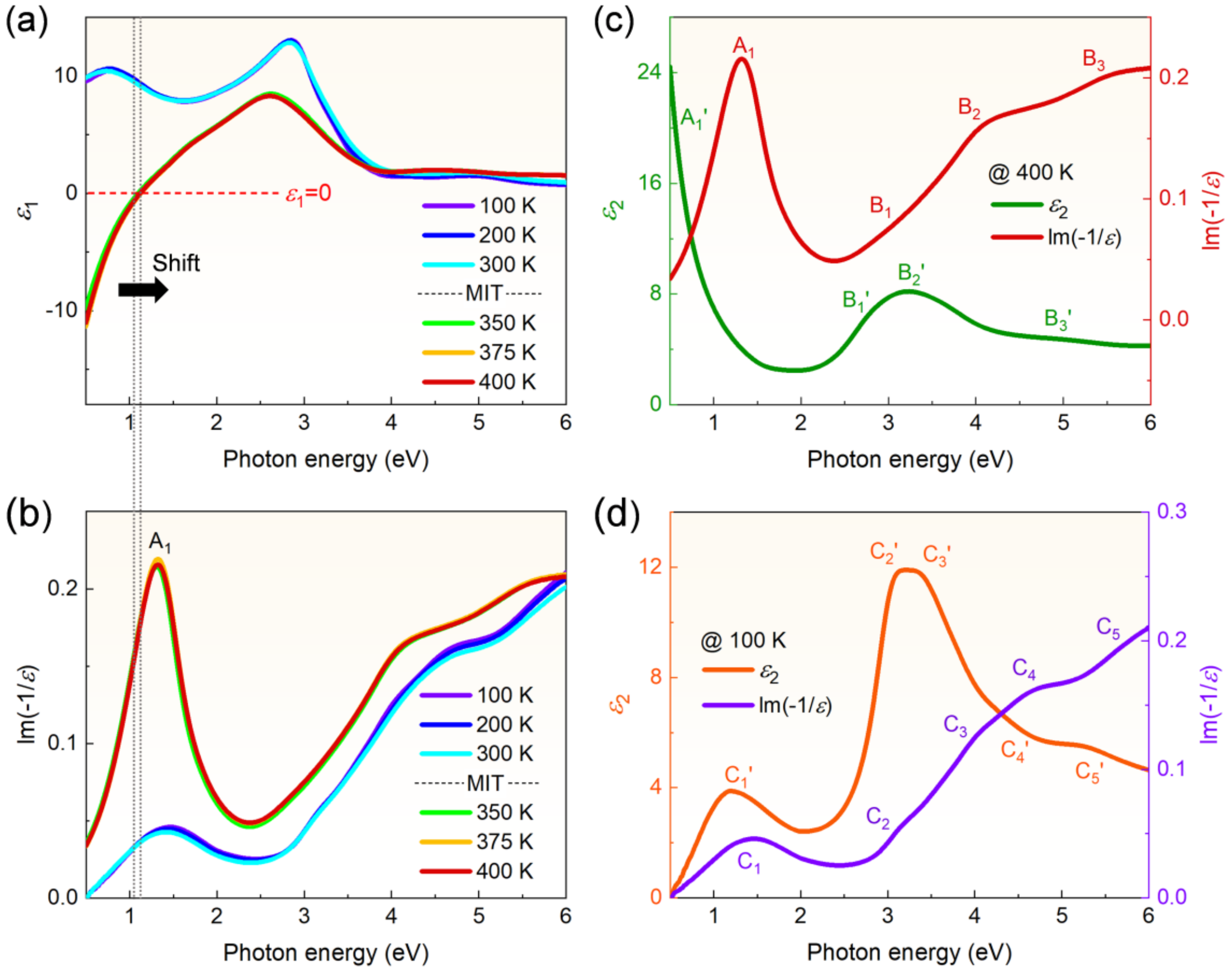

**Figure 5. The observation of plasmon excitations.** (a) Real component of the dielectric function, $\varepsilon_1$, at the respective temperatures above and below $T_{MIT}$ and, (b) shows the corresponding Loss Function (LF) spectra. (c-d) The $\varepsilon_2$ and LF spectra of $VO_2$ film at (c) 400 K and (d) 100 K.

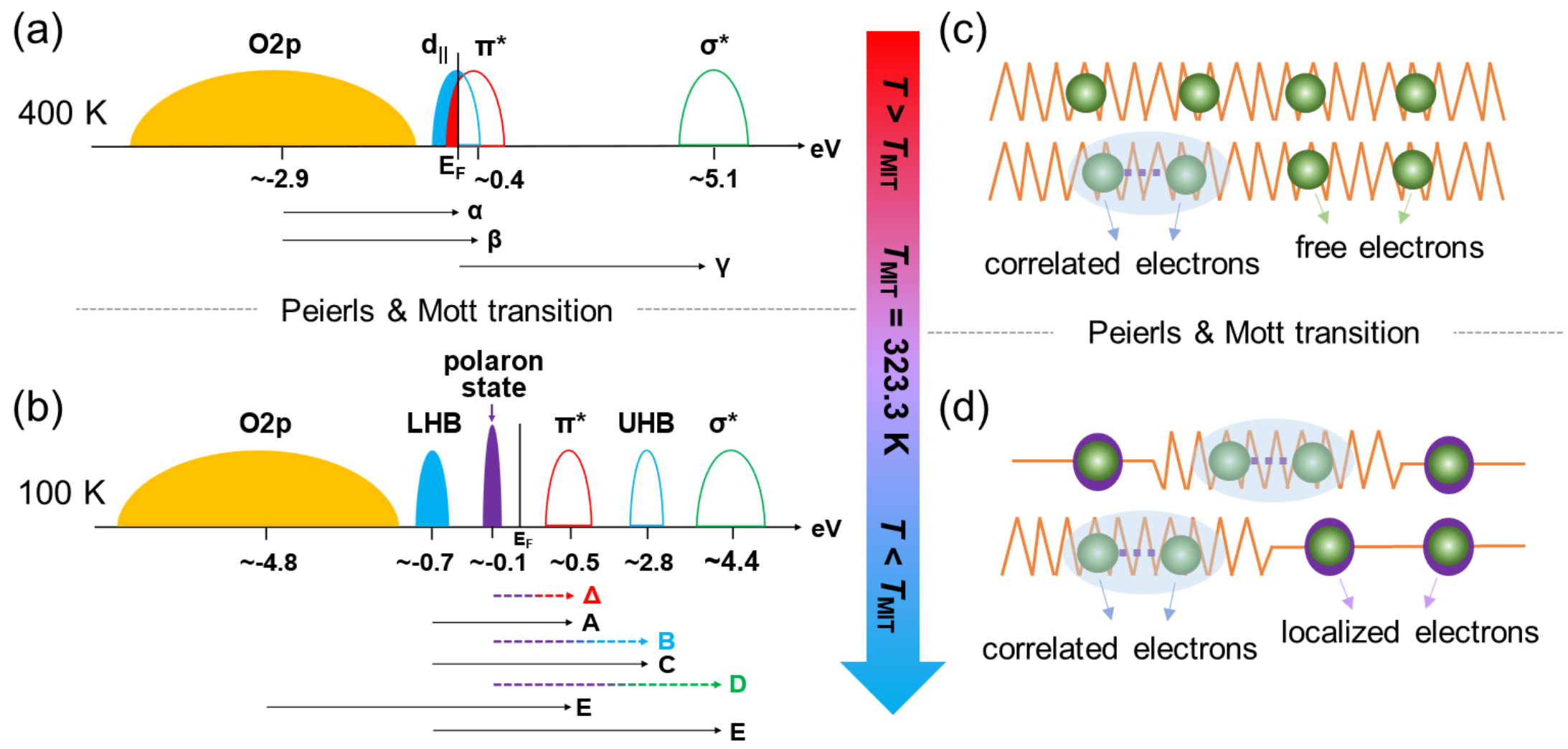

**Figure 6**. Changes to the electronic structures of the $VO_2/TiO_2$ film during the MIT process. Schematic models of the electron structures for $VO_2$ in (a) the metallic state at 400 K, and (b) in the insulating state at 100 K. The arrows and the respective letters denote the interband transitions in metallic and insulating phase $VO_2$ as registered by the optical measurements previously displayed in Figures 3a-b, respectively. (c-d) Schematics depicting how electron correlations evolve during the MIT transition process. (c) Electrons are delocalized in the metallic state where collective oscillations of free electrons and correlated electrons are present. (d) As the temperature decreases, the number of correlated electrons increases as an increasing number of electrons are localized. This results in the presence and coexistence of correlated electrons and small polarons in insulating phase $VO_2$.

# *Supplemental Information*

## Small polarons mediated near-room-temperature metal-insulator transition in vanadium dioxide and their hopping dynamics

*Xiongfang Liu[1,#], Tong Yang[2,#], Shanquan Chen[3,#], Jing Wu[4], Chi Sin Tang[1,5,*], Yuanjie Ning[1], Zuhuang Chen[3], Liang Dai[1], Mengxia Sun[1], Mingyao Chen[1], Kun Han[6], Difan Zhou[1], Shengwei Zeng[4], Shuo Sun[1], Sensen Li[7], Ming Yang[2,*], Mark B. H. Breese[5,8], Chuanbing Cai[1], Thirumalai Venkatesan[9], Andrew T. S. Wee[8,10], Xinmao Yin[1,*]*

[1]Shanghai Key Laboratory of High Temperature Superconductors, Institute for Quantum Science and Technology, Department of Physics, Shanghai University, Shanghai 200444, China

[2]Department of Applied Physics, The Hong Kong Polytechnic University, Kowloon, Hong Kong, 100872, China

[3]School of Materials Science and Engineering Harbin Institute of Technology, Shenzhen 518055, China

[4]Institute of Materials Research and Engineering (IMRE), Agency for Science, Technology and Research (A*STAR), 2 Fusionopolis Way, Innovis #08-03, Singapore 138634, Singapore

[5]Singapore Synchrotron Light Source (SSLS), National University of Singapore, Singapore 117603, Singapore

[6]Information Materials and Intelligent Sensing Laboratory of Anhui Province, Institutes of Physical Science and Information Technology, Anhui University, Hefei 230601, China

[7]Department of Electronic Engineering, School of Electronic Science and Engineering, Xiamen University, Xiamen 361005, China

[8]Department of Physics, Faculty of Science, National University of Singapore, Singapore 117542, Singapore

[9]Center for Quantum Research and Technology, University of Oklahoma, Norman, Oklahoma 73019, USA

[10]Centre for Advanced 2D Materials and Graphene Research, National University of Singapore, Singapore 117546, Singapore

[#]These authors contributed equally to this work.

[*]Corresponding author: slscst@nus.edu.sg (C.S.T.); kevin.m.yang@polyu.edu.hk (M.Y.); yinxinmao@shu.edu.cn (X.Y.)

# Content

# 1. Crystal Structure of Mott Insulator $VO_2$

## 1.1. Crystallographic Configuration of $VO_2$

$VO_2$ is a Mott insulator with reversible structural transition properties [1]. Close to the metal-insulator transition temperature ($T_{MIT}$) at or near room temperature, electronic phase transitions (the electrical resistance changes by several orders of magnitude between the metallic and insulating phases[2]) and structural transition (the dimerization of $V^{4+}$ ions along the $C_R$ axis direction, system from rutile phase to monoclinic phase[3]) occur almost simultaneously.

In the metallic phase above $T_{MIT}$, $VO_2$ has a rutile structure [4] (Figure. S1a). The neighboring V–V bond is ~ 2.85 Å arranged in chains along the $C_R$ axis. Each V-atom occupy the corner or center positions, and octahedrally coordinated with six oxygen ions. The space group of high-temperature $VO_2$ belongs is P42/mnm (136) [5]. In the insulating phase below $T_{MIT}$, the crystallographic configuration of $VO_2$ (Figure. S1b) is a distorted rutile structure, with the unit cell doubling compared to rutile type $VO_2$ [4]. Two different V–V distances of ~ 3.12 and ~ 2.65 Å between the nearest vanadium atoms form the zigzag atom chains [4] , The space group of insulating phase $VO_2$ is P21/C(14) [6]. The lattice structure of thin-film $VO_2$ is easily affected by the substrate [7]. Our $VO_2/TiO_2(001)$ film exhibits a tetragonal phase (main text Figure. 1a) in its insulating phase due to the epitaxial stress along the *a*/*b* plane [8].

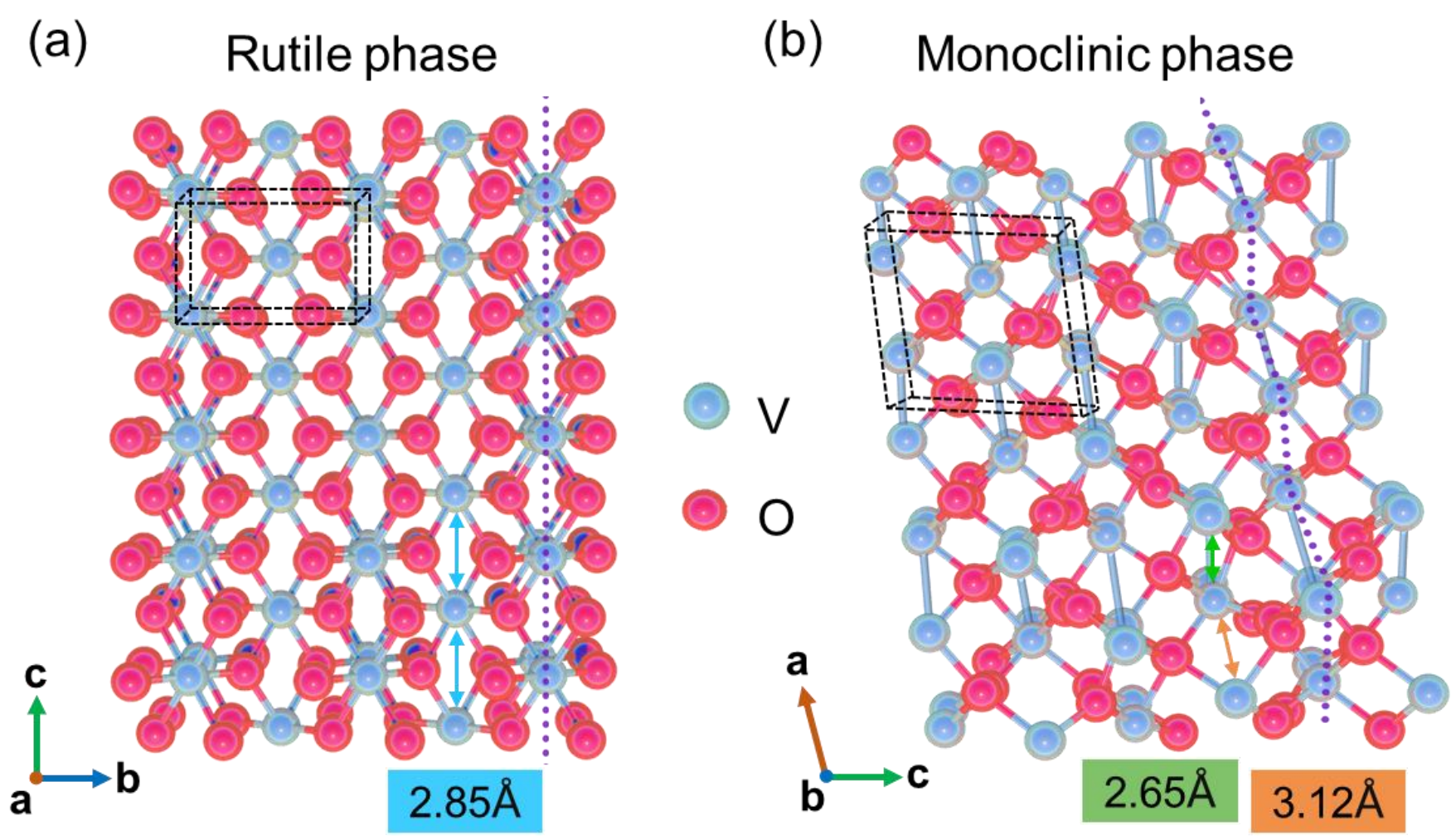

**Figure. S1**. Crystallographic configuration of the respective phases in $VO_2$ single crystal sample. (a) Metallic rutile phase. (b) Insulating monoclinic phase.

### 1.2 Sample Preparation

$VO_2$ films of thickness 28 nm were synthesized on $TiO_2$ (001) substrates by pulsed laser deposition (PLD). The rutile $TiO_2$ substrate is chosen as an ideal substrate for $VO_2$ film growth due to its stable thermal properties and similar lattice parameters. Deposition process of the $VO_2$ films on the $TiO_2$ (001) substrates take place at an optimized pressure and temperature of 10 mTorr and 673 ℃, respectively with a pulse laser repetition rate of 5 Hz and a laser fluence of ∼1.2 J/cm$^2$.

### 1.3 Crystalline Quality Characterization of $VO_2$ (High-resolution X-ray Diffraction, Electron Transport Measurements and Atomic Force Microscopy)

Vanadium oxides exist in multiple structural phases, tetravalent vanadium dioxide ($VO_2$) is the phases which can be readily oxidized to or reduced. $\theta$–$2\theta$ X-ray Diffraction (XRD) characterization is therefore conducted on the $VO_2/TiO_2$ film to determine their phase and crystallographic orientation. The diffraction peaks of sample are aligned with the standard tetragonal-like phase (tetragonal-like phase is a distorted by monoclinic phase [8]) of $VO_2$ and rutile phase of $TiO_2$, respectively. Subsequently, the Full Width at Half Magnitude (FWHM) obtained from the rocking curves (Figure. S2) can determine the crystalline quality of the $VO_2$ films. The FWHM data is fitted via using Gauss model, thus the FWHM of the fitted $VO_2$(002) reflection peak is 0.259 °.

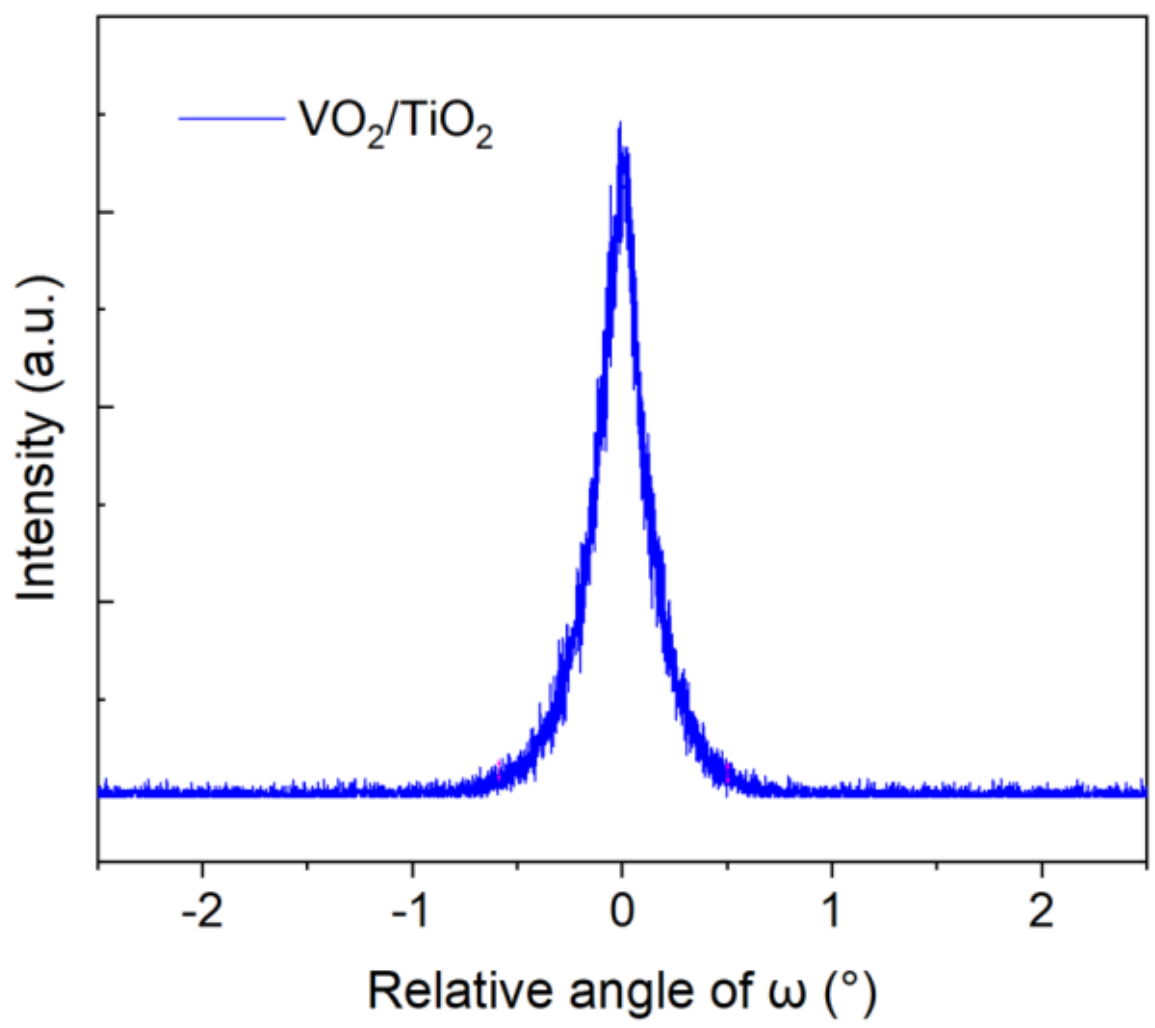

**Figure. S2.** X-ray rocking curve measurement on $VO_2/TiO_2$ film.

The electron transport data were taken using the standard four-probe method in a commercial Quantum Design physical property measurement system (PPMS). The transport measurements in our study focused on the metal-insulator transition properties of $VO_2$. The MIT temperature and temperature-dependent resistivity of the $VO_2$ film (main text Figure. 1c) are once again in very good agreement with transport measurements involving similar $VO_2$ film systems [7, 9-11]. The resistivity of $VO_2$ film is measured to be $5.9\times10^{-4}$ $\Omega\cdot$cm in the metallic phase at 400 K. The resistivity value increases by three orders of magnitude as temperature decreases into the insulating phase. The resistivity of $VO_2$ film is 9.9 $\Omega\cdot$cm at 250 K. The resistivity of $VO_2$ increases as temperature decreases. Given the measurement limitations of the instrument, the resistance of the $VO_2$ film may surpass the instrument's range during transport measurements conducted below ~211 K. As shown in the inset of Figure 1c, the MIT temperature ($T_{MIT}$) of $VO_2$ film during the heating cycle is ~323.3 K. Meanwhile, the MIT takes place at ~314.7 K during the cooling process with a small thermal hysteresis width of ~8.6 K. $VO_2/TiO_2$ sample with an abrupt one-step transition by four orders of magnitude further affirms that our $VO_2$ sample meet the optical experimental requirements. Generally, there is no obvious MIT phase transition in $VO_2$ films with large amount of oxygen vacancies (Figure S3).

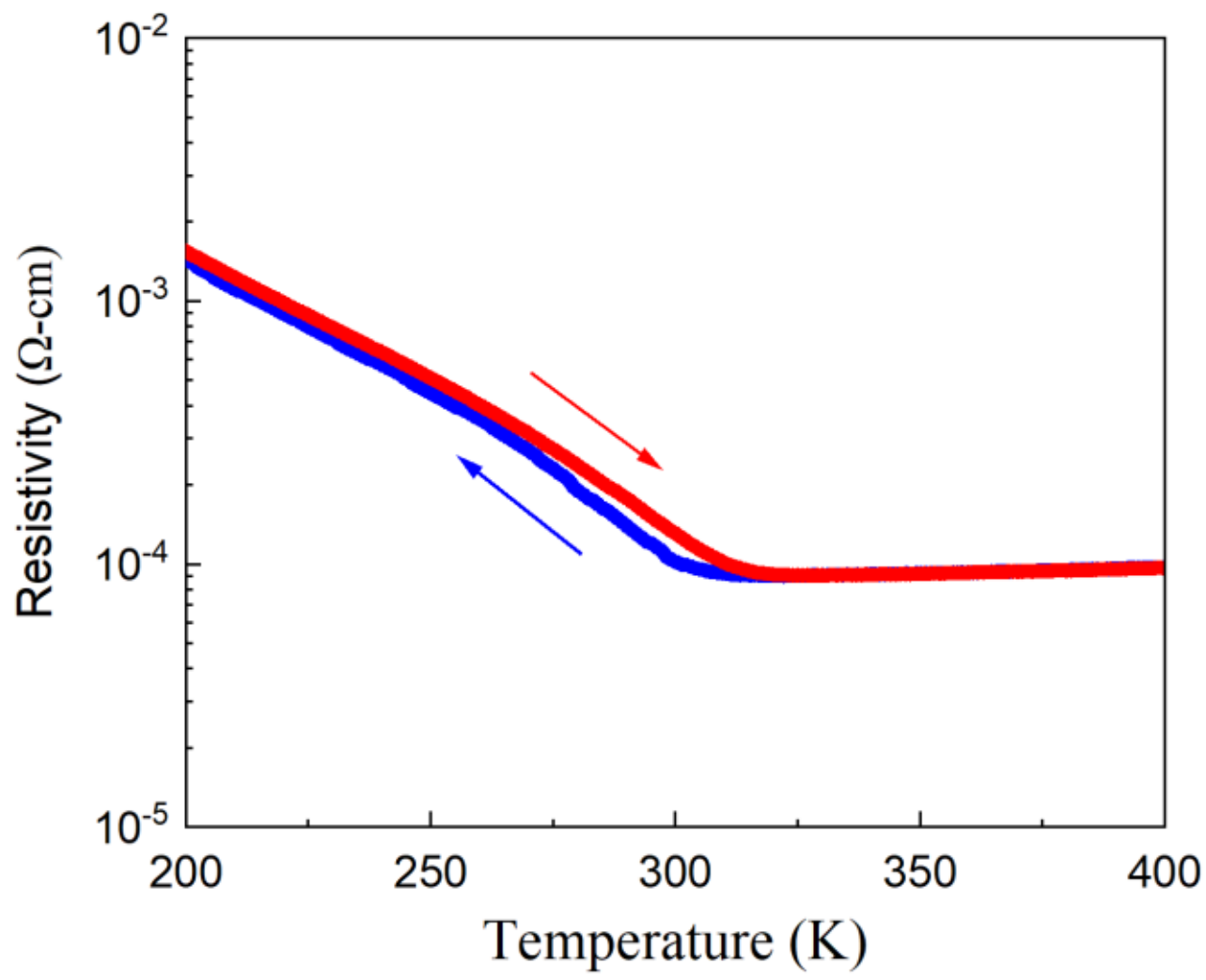

**Figure. S3**. Transport measurement of $VO_2$ film with large amount of oxygen vacancies.

The surface morphology of the $VO_2/TiO_2$ film was measured via atomic force microscopy (AFM). The surface roughness of the $VO_2/TiO_2$ film is 0.48 nm, which shows that our deposited $VO_2$ film has homogeneous grains and its surface is very flat and smooth (Figure S4).

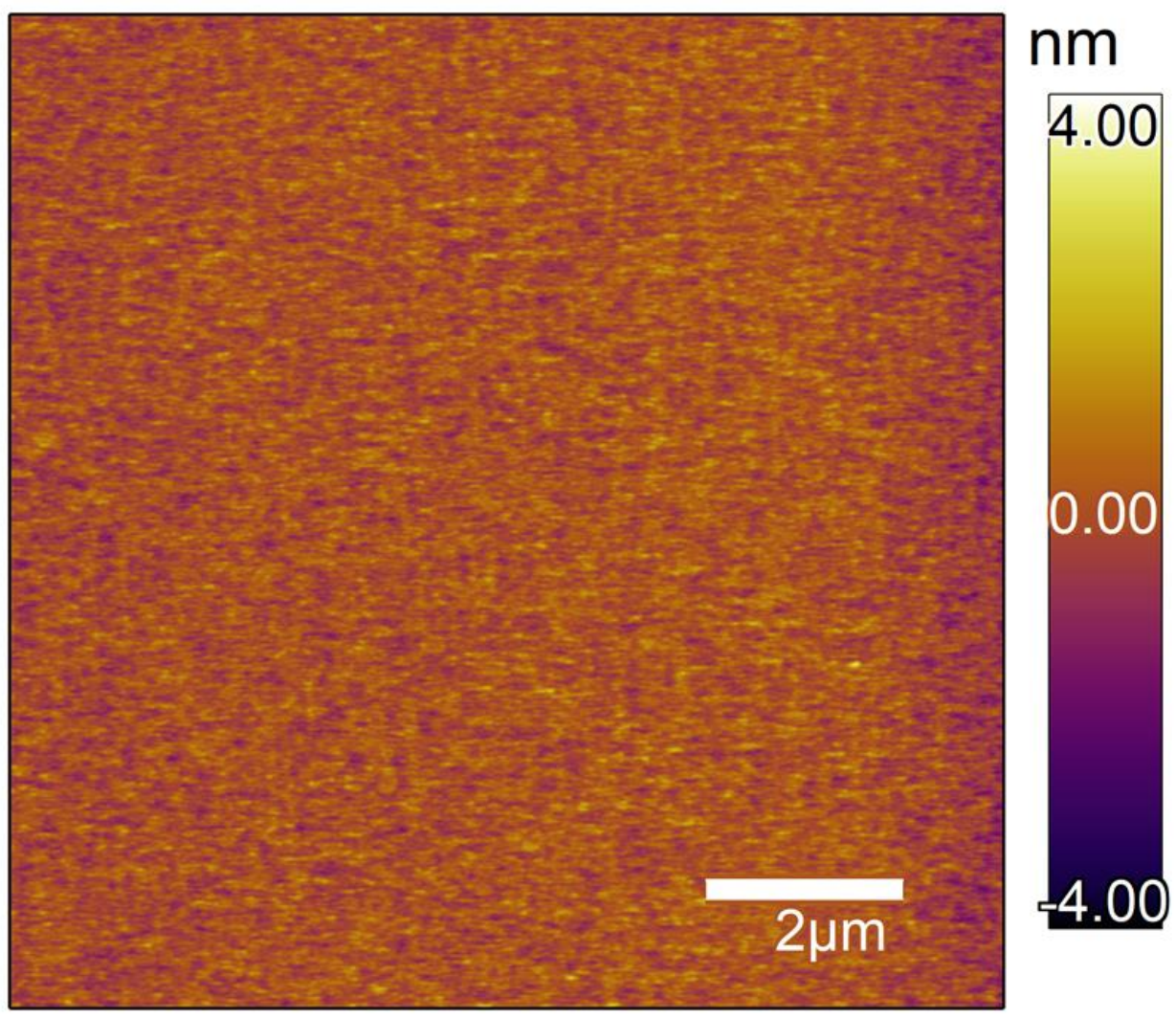

**Figure. S4**. AFM image of $VO_2$ film with a roughness of 0.48 nm.

## 2. Temperature-dependent Spectroscopic Ellipsometry Measurements

## 2.1. Experimental Principle and Methods

High-resolution spectroscopic ellipsometry (SE) technique offers extensive coverage across a wide optical energy range. Coupled with its capacity for temperature- and angle-dependent measurements, it proves to be a valuable technique for the identification and examination of quasi-particle excitations across a diverse array of materials [12-14]. Spectroscopic ellipsometry technique also proves highly beneficial as it enables the analysis of interactions between electrons and lattice, a phenomenon stemming from photon excitation, through the elucidation of optical spectra. With the above consideration, SE is a crucial tool to study the optical properties of films [15]. Particularly, SE is advantageous for its non-destruction, high sensitivity, and comprehensive response, suitable for the characterization of strongly-correlated system, such as $VO_2$ films [16, 17]. Without rely on a Kramers-Kronig transformation, SE can directly estimate the complex elements of the dielectric tensor from a single measurement. Using this method, SE can be extremely sensitive to charge distributions of surface properties [18]. The Schematic illustration of the spectroscopic ellipsometry experimental measurements as shown in Figure. S5.

Spectroscopic ellipsometry (SE) measurements are conducted using a custom-made J. A. Woollam Co., Inc Variable Angle Spectroscopic Ellipsometer (VASE) in the photon energy range of 0.5-6.0 eV at incident angle of 70° with respect to the plane normal. Measurements are conducted at temperatures between 100 K and 400K in a high-vacuum chamber at a base pressure of ~ $10^{-9}$ mTorr. SE measurements measure the ellipsometric parameters $\Psi$ (amplitude ratio between p- and s- polarized reflected light) and $\Delta$ (phase difference between of p- and s-polarized reflected light). The dielectric coefficients and optical conductivity of the $VO_2/TiO_2$ films are extracted from the ellipsometric parameters $\Psi$ and $\Delta$ utilizing an air/$VO_2$/$TiO_2$ multilayer model.

We measured the ellipsometric parameters $\Psi$ (amplitude ratio between p- and s-polarized reflected light) and $\Delta$ (phase difference between of p- and s-polarized reflected light). The dielectric coefficients and optical conductivity of the $VO_2/TiO_2$ films are extracted from the ellipsometric parameters Ψ and Δ utilizing an air/$VO_2/TiO_2$ multilayer model, where the $VO_2$ film consists of an average homogeneous and uniform medium.

Ellipsometric parameters Ψ and Δ are defined as

$$\tan\boldsymbol{\Psi}\ \boldsymbol{exp}(\mathbf{i\Delta}) \equiv \frac{\boldsymbol{r_p}}{\boldsymbol{r_s}} \qquad \textbf{(S1)}$$

Where $r_p$ and $r_s$ represent the reflectivity of p- and s- polarized light. Combined with the Fresnel equations, these variables are expressed as

$$\boldsymbol{r_p^{ij}} = \frac{\boldsymbol{n_j}\cos\boldsymbol{\theta_i} - \boldsymbol{n_i}\cos\boldsymbol{\theta_j}}{\boldsymbol{n_j}\cos\boldsymbol{\theta_i} + \boldsymbol{n_i}\cos\boldsymbol{\theta_j}} \qquad \textbf{(S2)}$$

and

$$\boldsymbol{r_s^{ij}} = \frac{\boldsymbol{n_j}\cos\boldsymbol{\theta_i} - \boldsymbol{n_i}\cos\boldsymbol{\theta_j}}{\boldsymbol{n_j}\cos\boldsymbol{\theta_i} + \boldsymbol{n_i}\cos\boldsymbol{\theta_j}} \qquad \textbf{(S3)}$$

Where n and θ is the refraction index and incident angles with respect to the plane normal respectively. Here, i and j are viewed as two materials. The complex dielectric function $\varepsilon(\omega) = \varepsilon_1(\omega) + i\varepsilon_2(\omega)$ of media can be obtained by

$$N^2 \equiv \varepsilon \qquad \text{(S3)}$$

where $\omega$ represent the photon frequency.

In typical two layers thin-film: $VO_2$ film on $TiO_2$ substrate, the reflectivity can be expressed as follows

$$r_{multi} = \frac{r_{vac,film} + r_{film,sub}exp\left(i2\delta_{film}\right)}{1 + r_{vac,film}\cdot r_{film,sub}exp\left(i2\delta_{film}\right)} \qquad \text{(S5)}$$

where

$$\boldsymbol{\delta_{film}} = \frac{\mathbf{2}\boldsymbol{\pi d_{film}}}{\boldsymbol{\lambda}}\boldsymbol{n_{film}}\cos\boldsymbol{\theta} \qquad \textbf{(S6)}$$

where the subscripts multi and vac represent $VO_2$ film on $TiO_2$ substrate multilayer system and the vacuum, respectively. $\delta_{film}$ is the change in light phase as it reflects from the $VO_2$ film, λ denotes the light wavelength, and $d_{film}$ is the thickness of the $VO_2$ film grown by PLD. As for the refractive index of the substrate, one clean bulk $TiO_2$ was measured by the spectroscopic ellipsometry.

After measuring complex dielectric function of $TiO_2$ substrate, the complex dielectric function of $VO_2$ films can be extracted from Ψ and Δ through fitting with Drude-Lorentz oscillators according to

$$\tilde{\varepsilon}(\omega) = \varepsilon_\infty + \sum_{k=1}^{n} \frac{{\omega_{pk}}^2}{{\omega_{0k}}^2 - \omega^2 - i\gamma_k\omega} \tag{S7}$$

$$\tilde{\varepsilon}(\omega) = \varepsilon_1(\omega) + i\varepsilon_2(\omega) \tag{S8}$$

where $\omega_0$ is the oscillator frequency, $\omega_p$ is the plasma frequency, $\gamma$ is the line-width. $\varepsilon_\infty$ is represent the higher-frequency oscillators.

Subsequently, the real part of the optical conductivity $\sigma_1$ can be calculated with the following equation[19]:

$$\sigma_1 = \omega \cdot \varepsilon_0 \cdot \varepsilon_2 = \frac{E}{\hbar} \cdot \varepsilon_0 \cdot \varepsilon_2 = \frac{2\pi\varepsilon_0}{h} \cdot E \cdot \varepsilon_2 \tag{S9}$$

where $\omega$ is the frequency of photon, $E$ is the energy of photon, $\varepsilon_2$ is the imaginary part of the dielectric function, $\varepsilon_0$ is a constant which is approximately equal to $8.85\times 10^{-12}$ F/M, $\hbar$ is also a constant, after unit conversion, the value is about $4.14\times10^{-15}$ eV·s.

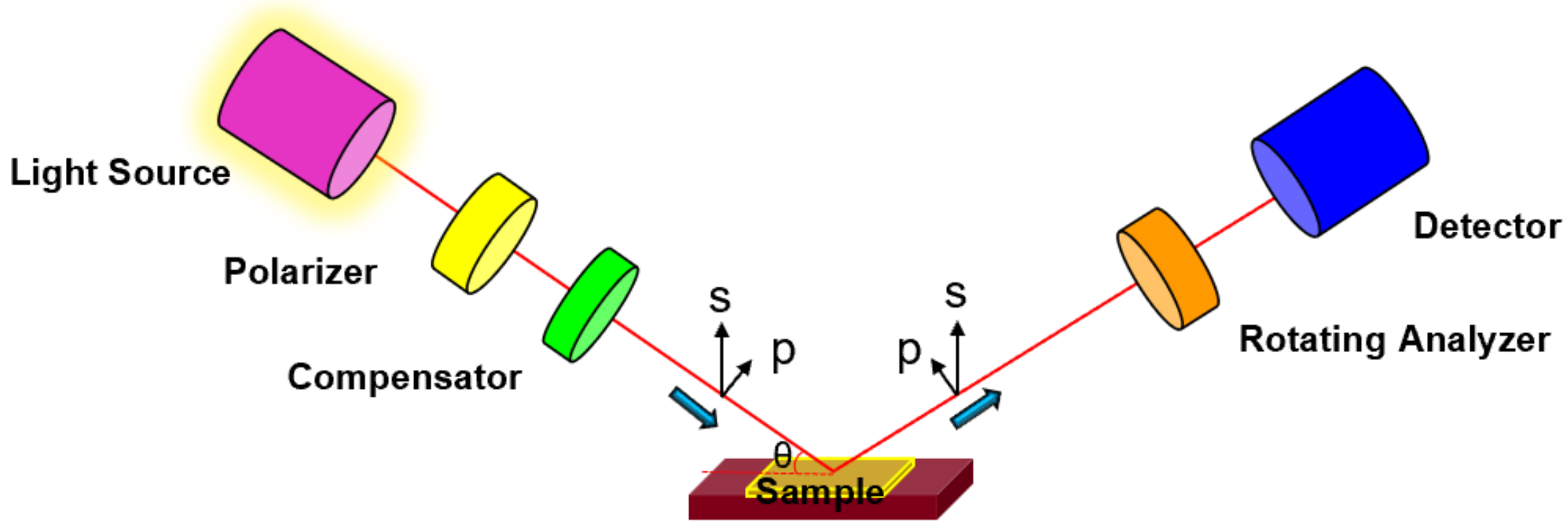

**Figure. S5.** Schematic of spectroscopic ellipsometry with the respective optical components.

## 2.2 Temperature-dependent Spectroscopic Ellipsometry Characterization

Spectroscopic ellipsometry represents a highly versatile tool for probing and effectively distinguishing between conventional optical band transitions and quasiparticle dynamics that arise due to many-body effects in strongly-correlated materials. It is thus widely utilized to establish the optical response and electron structures in a wide range of condensed matter systems and strongly-correlated materials at the nanoscale [12-14]. Detailed optical characterization of quasiparticles was conducted on $VO_2/TiO_2$ film using spectroscopic ellipsometry at temperatures between 100 K and 400 K where the MIT takes place. Optical responses of the $VO_2$ film were collected with incident photon energy between 0.5 and 6.0 eV. With the transport measurements confirming the MIT phenomena of the $VO_2$ film during the heating and cooling processes, it behaves as an insulator with a band gap (~ 0.6 eV) below 323.3 K ($T_{MIT}$) [20, 21]. It then transforms into a rutile metal at temperatures above 323.3 K [22]. These transport properties have been confirmed by the optical characterization using spectroscopic ellipsometry with prominent changes to the optical parameters caused by the transition process especially with regards to its energy bands and quasiparticle dynamics in both the metallic and insulating states. In the spectroscopic ellipsometry measurements, we observe the traces of three types of quasiparticles excitations, small polarons, metallic plasmons and correlated plasmons, respectively. The presence of polaron state related to small polarons is detected by analyzing the $\sigma_1$ absorption peaks in the insulating state. Meanwhile, metallic plasmons and correlated plasmons are found to distribute in the loss-function spectra.

### 2.3 The Number of Effective Carriers

The number of effective carriers per formula unit $[N_{eff}(\omega)]$ is as follows:

$$N_{eff}(\omega) = \frac{2m_0 V}{\pi e^2}\int_{w_2}^{w_1}\sigma_1(\omega')d\omega' \tag{S10}$$

As shown in Figure. S6, the reduction of $N_{eff}(\omega)$ in region I as temperature decreases is closely accompanied by the enhancement of $N_{eff}(\omega)$ in in regions II, III, IV, V. The changes of spectral weight and the number of the effective carriers are consistent with the conductivity behavior, indicating the occurrence of the MIT process. This is also an important sign of strong electronic correlations [23].

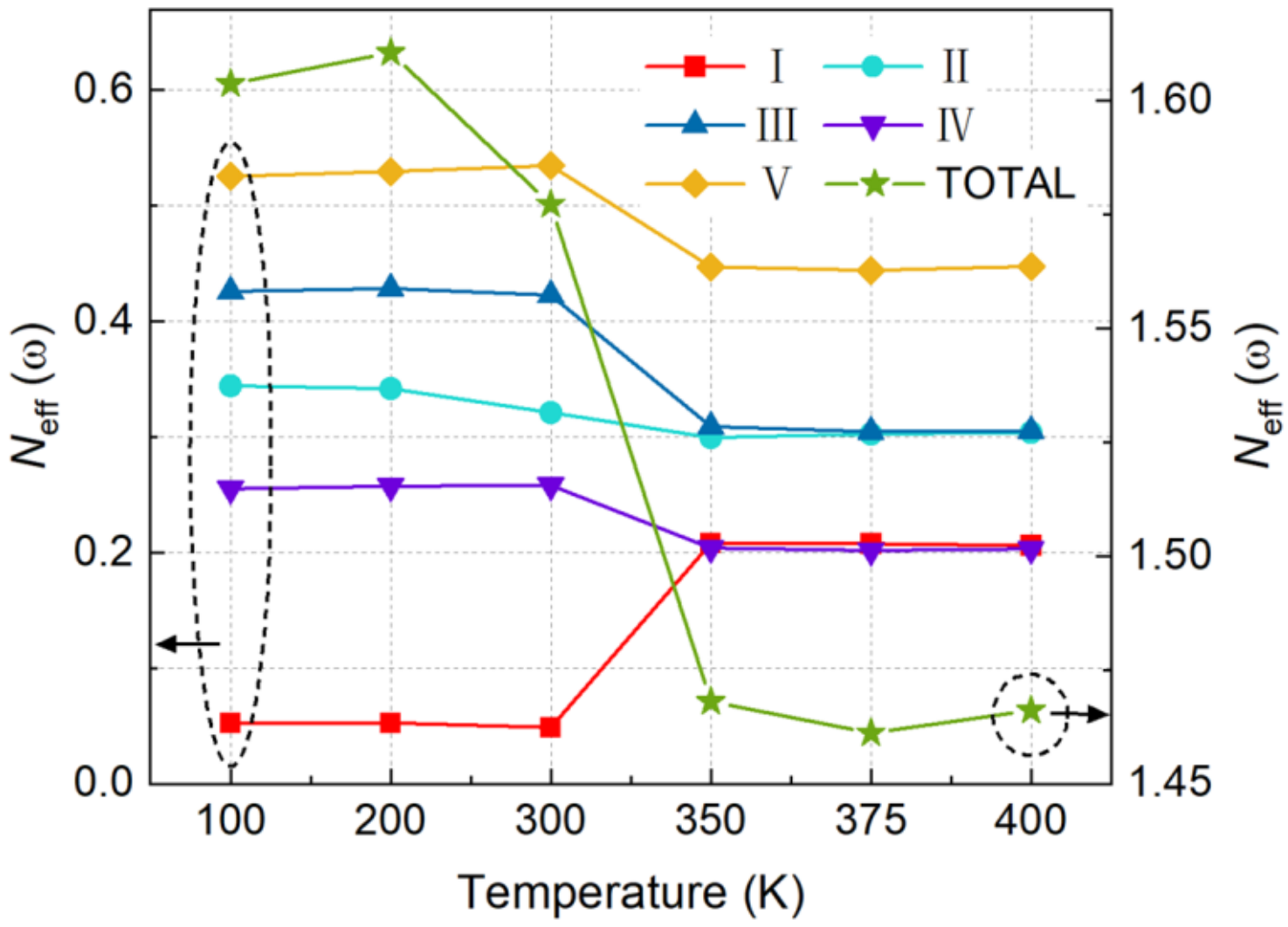

**Figure. S6.** The number of the effective carriers per formula unit $[N_{eff}(\omega)]$ as a function of temperature for different spectral regions: I (0-1.62 eV), II (1.62-3.28 eV), III (3.28-4.10 eV), IV (4.10-4.72 eV), (4.72-6.0 eV) and total (0-6.0 eV).

### 2.4. Drude-Lorentz Model

The dielectric functions of metals and semiconductors with high carrier concentrations exhibit strong free carrier absorption characteristics, which can be represented by the Drude model. The carrier concentration and mobility in metals/semiconductors can be estimated using the Drude model for analysis. However, there are two contributions in both $\varepsilon_1$ and $\varepsilon_2$ spectrums of metals/semiconductors in reality, free-carriers and bound electrons. Therefore, the dielectric function of a metal/semiconductor is expressed as

$$\tilde{\varepsilon}(\omega) = \varepsilon_1(\omega) + \varepsilon_2(\omega) \tag{S11}$$

$$\tilde{\varepsilon}(\omega) = \tilde{\varepsilon}_{free}(\omega) + \tilde{\varepsilon}_{bound}(\omega) \tag{S12}$$

The free electron component as represented by $\tilde{\varepsilon}_{free}$ is modelled by the Drude model while the bound electrons as represented by $\tilde{\varepsilon}_{bound}$ by the Lorentz model. This component is introduced to analyze the relationship between the dielectric polarization phenomenon and the applied electromagnetic field to assist in modelling the dielectric property of the system.

Hence, the Drude-Lorentz model has been used for in modelling the dielectric functions of several semiconducting systems which possess both free and bound electronic properties. It easily models the dielectric properties of such systems over a wide spectral range and provide the perfect Gaussian line shape for the construction of optical curve fitting [24].

## 2.5 Optical Transitions Analysis in Optical Conductivity Spectra

In the metallic phase at 400 K, anti-ferroelectric displacement along the [110] direction decreases the *p*-*d* hybridization, causing the $\pi^*$ band to shift downward towards the Fermi level [5, 25]. As a result, there are two half-filled bands, $d_{||}$ and $\pi^*$ located at the Fermi level in the metallic state. Three features in the $\sigma_1$ spectra can be attributed to optical transitions between occupied energy bands and unoccupied energy bands [26]. Combining with our analysis of spectroscopic ellipsometry, feature α (~ 2.75 eV) can be attributed to the optical transition from the filled O *2p* band to the half-filled $d_{||}$ band, while a weak shoulder feature β (~ 3.38 eV) is ascribed to the optical transition from the filled O *2p* band to the half-filled $\pi^*$ band. Meanwhile, feature γ (~ 5.07 eV) is attributed to the optical transition between half-filled $d_{||}$ band and empty $\sigma^*$ band.

Abrupt MIT occurs at 323.3 K, leading to significant changes in both the electrical conductivity and lattice structure. We aim to account for these optical conductivity features by comparing our results to earlier theoretical and experimental studies [27, 28]. In the insulating phase at 100 K, previously studies revealed that the energy band of $VO_2$ evolves with temperature [29], specifically, the energy bands near the Fermi level shift closer to it with increasing temperature. This reduces the energy difference between the minimum conduction band and maximum valence band. The optical conductivity data of insulating $VO_2$ film shows the feature labelled A is situated at ~ 1.26 eV at 100 K which redshifts to ~ 1.16 eV at 300 K. This feature can be attributed to the optical transition between adjacent energy bands that near the Fermi level that is commonly reported in the insulating state [30, 31].

The Lower Hubbard band (LHB) and Upper Hubbard band (UHB) are two subbands split by half-filled Hubbard model due to the influence of Hubbard intra-atomic Coulomb repulsion energy [26]. LHB is a full-filled energy band which located below the Fermi level while the UHB is an empty energy band which located above the Fermi level and near the $\pi^*$ bands [29]. Hence, feature C at ~ 3.39 eV in the insulating phase can be attributed to the interband transition between the LHB and UHB [25, 31], revealing strong Peierls charge-density wave ordering from dimerization in insulating $VO_2$ [17, 32]. In the energy regions away from the Fermi level. The hump feature E at ~ 5.33 eV with high-intensity spectral weight in insulating state can be defined as the transition from the O *2p* bands to the empty $\pi^*$ band and from LHB to empty $\sigma^*$ band [30, 33].

## 2.6 Plotting the Differential Spectra of the Optical Conductivity

A more in-depth analysis of the spectroscopic ellipsometry data is conducted and they are plotted in the form of a differential spectra of optical conductivity between various temperatures and the 300 K in insulating state, $\Delta\sigma_1(\omega, T)=\sigma_1(\omega, T)- \sigma_1(\omega, 300 K)$, as shown in Figure S7. We now focus on the low-energy region (0.5 eV—2.0 eV). An independent feature can be observed at ~0.7 eV across different temperatures. This feature may be attributed to the optical transition from the polaron state to the $\pi^*$ band.

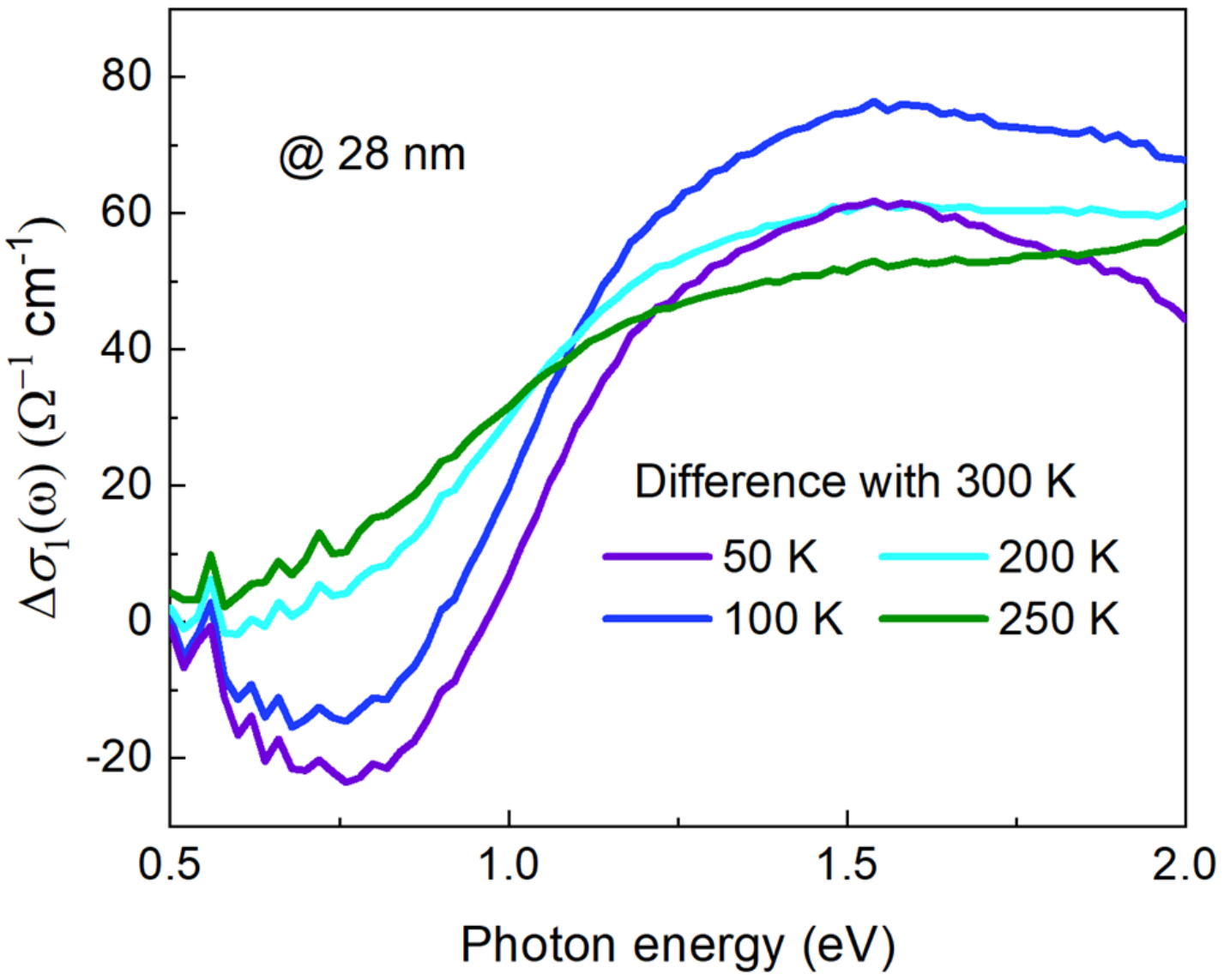

Figure S7. The differential spectra of the optical conductivity for 28nm-$VO_2$/$TiO_2$ film.

The same situation is also observed in the 27.0 nm-$VO_2$/$TiO_2$ films. A set of optical conductivity and differential spectra of the 27-nm $VO_2$ film is displayed in Figure S8. At the lower energy position of ~0.7 eV, an optical transition feature is also observed and it can also be attributed to the emergence of the polaron state.

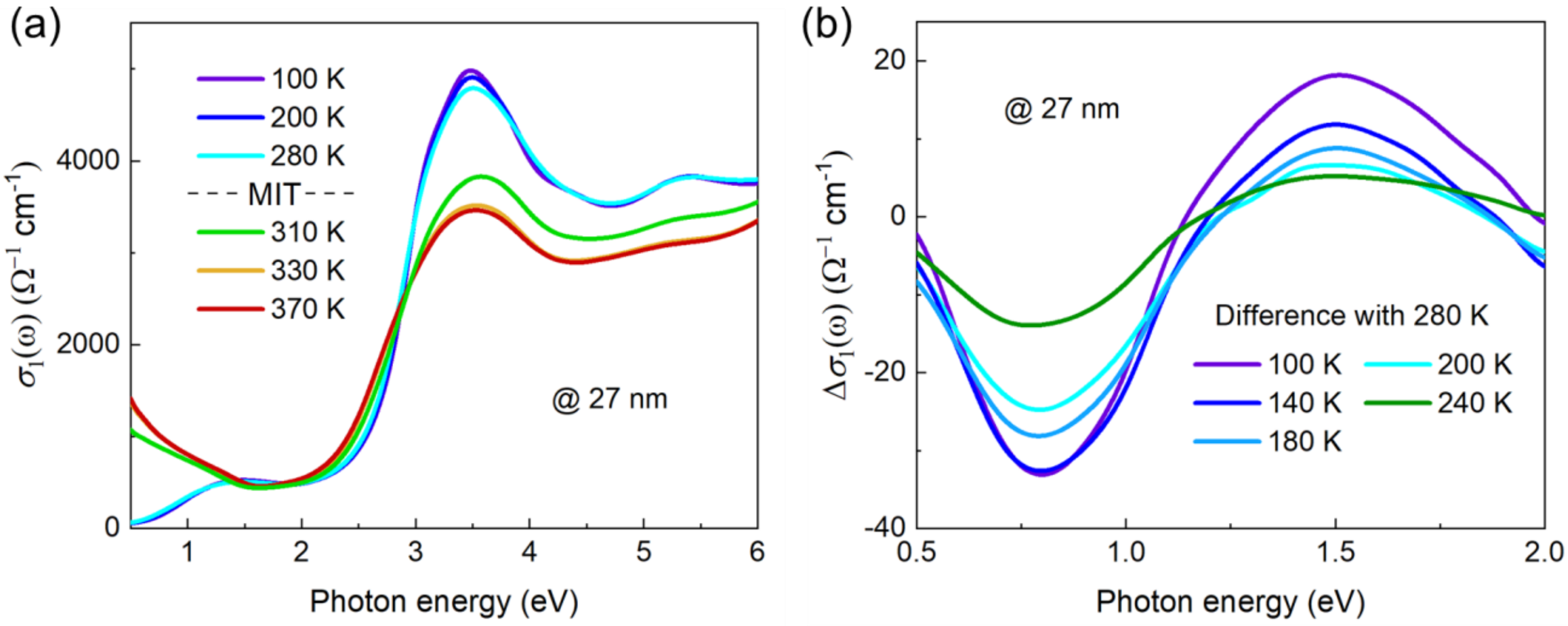

Figure S8. (a) $\sigma_1$ spectra of 27nm-$VO_2$/$TiO_2$(001) film from 0.5 to 6.0 eV at the respective temperatures in both its insulating and metallic phases. (b) Differential spectra of optical conductivity for 27nm-$VO_2$/$TiO_2$ film.

## 2.7 Dynamical Response Loss-Function

Further insights to the loss-function(LF) between the metallic and insulating vanadium dioxide systems which can successfully describe the difference between the electron behaviors of $VO_2$ around $T_{MIT}$ [34]. In a broader sense, the dynamic energy loss measurement technique can lay a foundation for the analysis of the optical properties in $VO_2$ and other phase change materials [35, 36].

The optical LF is further derived with the spectra from the dielectric function based on the following expression:

$$LF = Im\left(-\frac{1}{\tilde{\varepsilon}}\right) = \frac{\varepsilon_2}{{\varepsilon_1}^2 + {\varepsilon_2}^2} \tag{S13}$$

where $\varepsilon$ denotes the complex dielectric function ($\tilde{\varepsilon} = \varepsilon_1 + i\varepsilon_2$).

## 2.8 The Observation of Plasmon Excitations

To further analyze the quasiparticle features and opto-electronic properties of the $VO_2$ system and distinguish between the insulating and metallic phases, the optical loss-function (LF) is further derived with the spectra displayed in main text Figures. 3(g-h). Metallic phase $VO_2$ shows clear metallic behavior with the prominent Drude response in $\sigma_1$ (main text Figure. 2a) and zero-crossing of $\varepsilon_1$ (main text Figure. 3g) at ~ 1.10 eV at 350 K. This zero-crossing blueshifts monotonically to ~ 1.12 eV at 400 K – near the energy position of loss-function peak $A_1$ (main text Figure. 3h) marking the presence of metallic plasmons (peak $A_1$) attributed to the collective excitation of free charges [35]. While the coincidence of the zero-crossing in the $\varepsilon_1$ spectra and the LF peak $A_1$ at the same photon energy position marks the presence of metallic plasmons in metallic phase $VO_2$, note the very slight disparity in the respective energy position as shown in Figures. 3(g-h) where it can be attributed to the presence of free electron scattering in the $VO_2$ crystalline system [37]. Since $VO_2$ exhibits metallic behavior at temperatures above $T_{MIT}$, metallic plasmons can also be detected. Our detection of metallic plasmons is in complete agreement with previous literature report [38-40]. Figure. S9 shows the peak positions of metallic plasmons (peak $A_1$) present obvious temperature-dependent properties. This is related to the temperature-dependence of concentration of charge carriers in the metallic state [41].

Correlated plasmons present in strongly-correlated systems have been reported in a diverse class of materials [42-44]. $VO_2$ is a quintessential strongly-correlated system, using methods similar to those employed in previous studies [42, 45], we have successfully detected correlated plasmons in $VO_2$. An essential observation is depicted in the $\varepsilon_2$ and LF spectra of $VO_2$ [main text Figures. 3(i-j)]. An essential observation is depicted in the $\varepsilon_2$ and LF spectra of $VO_2$ (Figures 3i,j). In the metallic phases (400 K), apart from the Drude peaks ($A_1$), several features, namely $B_1$, $B_2$ and $B_3$ can be observed in the LF spectra. Notably, the energy positions of features $B_1$, $B_2$ and $B_3$ in the LF spectra closely align with those of another group of three features in the $\varepsilon_2$ spectra: $B_1'$, $B_2'$ and $B_3'$, respectively. In the insulating state (100 K) of LF, several features, namely $C_1$, $C_2$, $C_3$, $C_4$ and $C_5$ also correspond one-to-one with the features in $\varepsilon_2$ (features $C_1'$, $C_2'$, $C_3'$, $C_4'$ and $C_5'$). These observations suggest that the simultaneous presence in both the $\varepsilon_2$ and LF spectra where they indicate a coupling between optical and plasmonic excitations in metallic and insulating $VO_2$ [42] (Supplementary, Section 2.7). Therefore, features $B_1$, $B_2$ and $B_3$ in the metallic phase and $C_1$, $C_2$, $C_3$, $C_4$ and $C_5$ in the insulating phase $VO_2$ can be regarded as a form of correlated plasmons, arise from the collective excitations of correlated electrons.

The observation and characterization of both conventional (metallic) and correlated plasmons in the distinct phases of the $VO_2$ film offer valuable insights into the evolution of collective charge dynamics within the system as it undergoes transformations between structural and electronic phases. Furthermore, a comprehensive understanding of the plasmonic properties in $VO_2$ is crucial for their utilization in optoelectronic devices.

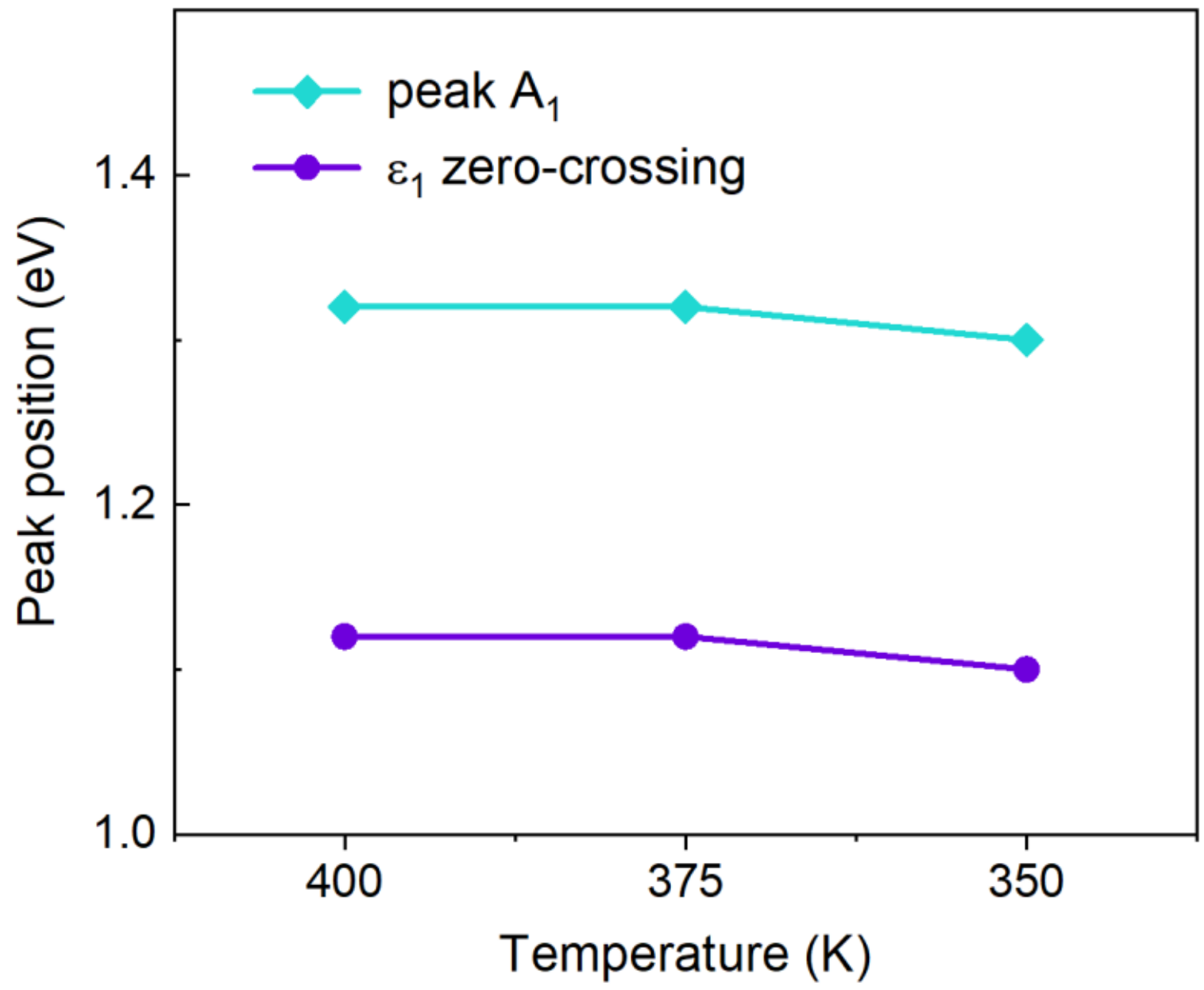

**Figure. S9.** The slight disparity of the position relationship between zero-crossings of $\varepsilon_1$ and peak $A_1$ can be attributed to electron scattering in the system.

## 3. Small Polarons Analysis

### 3.1 The Characteristics of Small Polarons

In a multidimensional system, the discernment between large polarons and small polarons serves as a crucial aspect in the realm of quasiparticles. These two distinct types of quasiparticles exhibit dissimilar effects on electron structures and optical transitions [46-48]. Large polarons, in particular, tend to induce a significant renormalization of the band gap, whereas small polarons, being more extensively spread across adjacent unit cells, may introduce discrete mid-gap states [49]. In light of our findings regarding the insulating $\sigma_1$ spectra (main text Figure. 2e), only a solitary mid-gap state emerges. Hence, we contend that the polarons observed in the $VO_2$ system under investigation can be classified as small polarons.

The localized interaction between excess electrons and the V lattices is responsible for the emergence of short-range effects. The local lattice undergoes relaxation in the vicinity of the V site, thus companying the creation of small polarons within the insulating state [50, 51]. These small polarons, propelled by the short-range electron-lattice interaction [52, 53], trigger electron transitions from a self-trapped ground state to adjacent localized states [50, 51, 54, 55]. Moreover, the presence of small polarons disrupts the magnetic moment and leads to an elongation of the V-O bond length, thereby imbuing the insulating phase of $VO_2$ with spatial anisotropy and orbital polarization [56]. Therefore, our investigation signifies that the formation of small polarons in $VO_2$ originates from the interplay between electrons and the lattices, giving rise to electron polaron state in close proximity to the valence band.

Another important discovery to note is that $V^{4+}$-$V^{4+}$ dimerized pairs are formed in its insulating state [55, 57-59] where the $V^{4+}$ oxidation state is an essential condition for the formation of small polarons in $VO_2$ [60, 61]. Meanwhile, the presence of $V^{3+}$ and $V^{5+}$states in metallic state $VO_2$ does not favor the formation of small polarons [62]. This accounts for the presence of small polarons in insulating state $VO_2$ while it is absent from its metallic phase, which is consistent with both the optical and transport analyses of the main text.

### 3.2 Small Polaron Hopping Model

The electrical conductivity of the system exhibits a linear temperature-dependence, as indicated by the relationship $\ln(\sigma T^{3/2}) \sim T$. This behavior takes into account the impact of lattice vibrations on the resonance integral [63, 64]. To describe the contribution of hopping to the electrical conductivity, the Small polaron hopping model proposed by Bryksin [53] is employed.

The hopping conductivity demonstrates a temperature dependence that can be expressed by the following relation:

$$\ln\left(\sigma T^{3/2}\right) = A + \frac{k_B T}{\varepsilon} - \frac{E_a}{k_B T} \tag{S14}$$

where A is the constant that independent of temperature, $K_B$ is the Boltzmann constant, ε is the vibrational energy scales and $E_a$ is the small polarons hopping activation energy.

The electronic transport mechanism observed in the $VO_2$ system suggests that the thermal lattice vibrations' magnitude influences the ability of small polarons to hop between lattice sites, thereby affecting the system's electrical conductivity [63]. In the low-temperature phase, the conductivity primarily arises from the hopping of small polarons. Notably, the experimental data have been fitted to match the curvature that is in line with the theoretical and experimental results below $T_{MIT}$. An activation energy ($E_a$) for small polaron hopping of $2.636\times10^{-2}$ eV can be derived, along with a lattice vibrational energy (ε) of $1.95\times10^{-3}$ eV, consistent with earlier investigations of $VO_2$ [63, 64]. However, a significant deviation between the hopping conductivity of small polarons and the experimental conductivity arises just before and during the onset of the MIT. The $\ln(\sigma T^{3/2})$ relationship in the metallic regime exhibits a substantially steeper rise compared to the expectations from the small polaron hopping model.

### 3.3 Small Polaron State

Having provided experimental evidence that small polarons are present in the insulating state $VO_2$ based on a detailed analysis of the transport data below and above the metal-insulator transition temperature, it is essential to clearly attribute features B and D observed in the optical spectra (Figure. 2e) to the presence of the polaron state in insulating phase $VO_2$. A previous theoretical study has indicated that polaron state is formed at energy region in the order of ~1 eV below the conduction band minimum (CBM) in large bandgap oxide systems [49]. With the relatively small bandgap of insulating $VO_2$ of ~ 0.67 eV, not only is the formation of the small polaron state at 0.6 eV below the empty $\pi^*$ band at 100 K (main text Figure. 4b) theoretically consistent, but it also agrees with experimental detected polaron state in other small bandgap oxide system [46].

### 3.4 Analysis for The Band Structures

As previously discussed, feature B in the optical conductivity spectra has been associated with the optical transition from the polaron state to the Upper Hubbard band (UHB), while feature D corresponds to the optical transition from the polaron state to the $\sigma$* band. Interestingly, the energy positions of both features B and D in the optical conductivity spectra, as shown in Figures. S10a-b, exhibit no significant changes within the temperature range from 100 K to 300 K. This suggests that there is also a corresponding position shift within the temperature changing in the polaron band along with the UHB and $\sigma^*$ band. Hence, the relative energy differences between these bands remains unchanged as seen in the positions of features B and D.

In order to analyze the unoccupied V-*3d* conduction bands in insulating $VO_2$ film at temperatures below the MIT ($T<T_{MIT}$), X-ray Absorption Spectroscopy (XAS) was performed on the O-*K* edge [65, 66]. Figure. S10c presents the XAS results from Ref [67], which exhibit two prominent maxima at ~ 529.5 eV and ~ 532.0 eV. These peaks can be attributed to the transition from the O *1s* state to the hybridized O *2p* state with the unoccupied conduction bands $\pi^*$ and $\sigma^*$, respectively[68]. Notably, the feature observed at ~ 530.5 eV is only present below $T_{MIT}$ [67] and can be considered as the UHB feature.

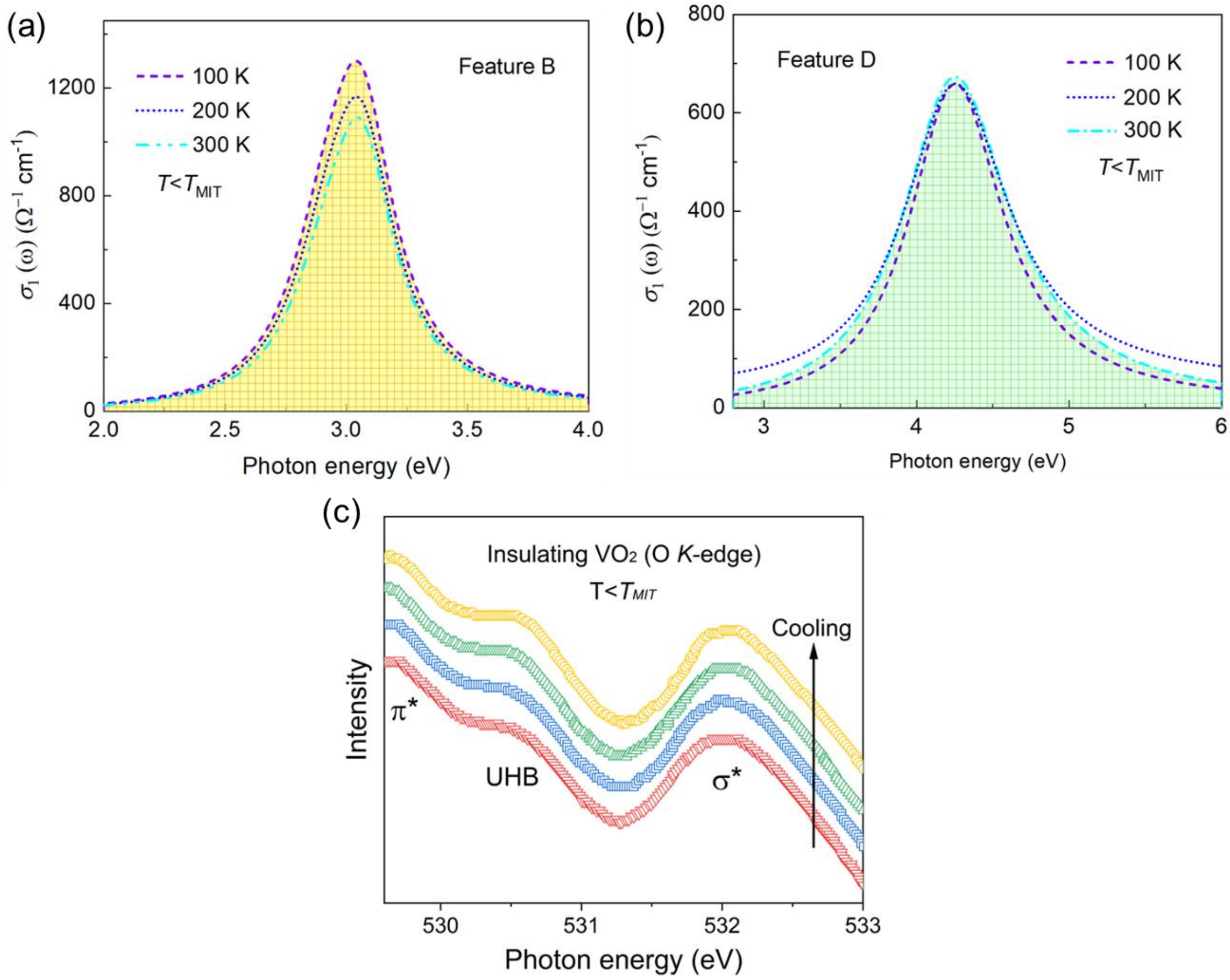

**Figure. S10.** The characteristics of peak B & D and O-*K* edge spectra elucidated from XAS measurements. (a-b) The optical conductivity of fitting features B and D (always located at ~ 3.02 eV and ~ 4.26 eV, respectively) in the insulating state at the temperature range from 300 K to 100 K. (c) Temperature dependence analysis of O-*K* edge in the XAS spectra is measured from Ref[67].

## 4. First-principles Calculations

Hence, by comparing evidences from our optical studies with previous theoretical studies detailing the presence of the polaron state in insulating phase $VO_2$, they provide substantial evidences that a small polaron state is present below the Fermi level with features B (~3.39 eV) and D (~4.26 eV) in the $\sigma_1$ spectra (main text Figure 2e) being the optical transitions from the polaron state to the UHB and $\sigma^*$ band, respectively. While our experimental studies have provided very conclusive evidences pointing to the presence of small polarons in insulating phase $VO_2$, further confirmation of this quantum quasiparticle will be made via a set of computational study as discussed thereafter.

Density functional theory (DFT) calculations were conducted using the Vienna ab initio Simulation Package (VASP) with the projector augmented-wave method. The Perdew-Burke- Ernzerhof (PBE) parameterized Generalized Gradient Approximation (GGA) has been adopted for the ion-electron interaction and the exchange-correlation interaction, respectively [69-72]. A plane wave basis with a kinetic energy cutoff of 600 eV was used to expand the electronic wavefunction. The valence states considered are O $2s^2 2p^4$ and V $3s^2 3p^6 3d^4 4s^1$. The $\Gamma$-centered $k$-point grid was set to 11×11×9 and 5×5×4 for the $VO_2$ unit cell and the 2×2×2 supercell, respectively. The total energy convergence criterion is $1.0\times10^{-5}$ eV and the structural relaxation was conducted until the force on each atom is less than 0.01 eV/Å. A Hubbard $U$ value of 5.3 eV has been applied to the V 3$d$ electrons to reproduce the experimental band gap of the insulating phase of $VO_2$ [73]. The 2×2×2 supercell containing 96 atoms was used to simulate the (non-)polaronic state. For the polaronic state simulation, in particular, one vanadium site was slightly displaced from its equilibrium position in order to break the symmetry constraint, whereas there is no such an initial displacement for the non-polaronic state simulation. This method has been widely used to study polarons in various oxide materials [74-77]. In particular, the value for Hubbard $U$ of 5.3 eV on the V3$d$ electrons adopted has been carefully selected in order to accurately describe the property of interest and the experimental band gap of the insulating $VO_2$ [77-79]. With the focus on the electronic structure of insulating $VO_2$ and its metal-insulator transitional properties, the value of $U$ has been fitted to reproduce the band gap size of the insulating $VO_2$ (Experiment: ~0.67 eV (100 K); DFT+U: 0.601 eV). With the ability of the DFT+U approach to accurately describe the polaron state and provide important insights to the experimental observations in strongly-correlated systems [48, 80-82], the consistent results between the band structure calculations and the experimental findings and that they agree with previous studies are clear indications that the observation of small polarons in insulating phase $VO_2$ is reliable and accurate. Moreover, it is noted that, to the best of our knowledge, neither the standard DFT method nor its advanced derivative methods can simultaneously and properly describe the magnetic and electronic properties of strongly correlated $VO_2$ systems [83]. We also note that no magnetism was experimentally observed in our samples. Therefore, the spin degree of freedom has not been considered throughout the DFT calculations. This is to circumvent the spurious effect induced by magnetism. As such, two electrons were added to the 2×2×2 supercell during the (non-)polaronic state simulation, namely, one electron for each degenerate spin channel. Similar approaches have been recently adopted to investigate other phases of vanadium oxides [77].

Meanwhile, these localized electrons can also be found in neighboring vanadium sites in the vicinity where the polaron states are formed (37). This is a clear indication that it is energetically more favorable for the excess electrons to be localized in the V lattice sites to form small polarons than in the non-polaron state where the electrons are homogeneously distributed across the lattice system. As compared to the non-polaron state counterpart, while the charge trapping at the V lattice site further increases the long V-V distance, the short V-V distance remains virtually unchanged. This is likely due to the aforementioned charge transfer from that neighboring V site. The calculated band gap is 0.601 eV, in agreement with that previously reported for the insulating phase of $VO_2$ [20, 21].

This polaron state emerged below the Fermi level in the polaron state and it mainly consists of the V$t_{2g}$ states. Very importantly, the results of these DFT studies show clear consistency with our experimental findings. Specifically, the mid-gap state that emerged in the visualized PDOS can be attributed to the formation of the polaron state. This provides compelling evidence to substantiate our experimental results for the presence of small polarons in insulating state $VO_2$.